\documentclass{article}
\usepackage{graphicx}
\usepackage{epsfig}
\usepackage{amsfonts}
 \usepackage{amssymb}
\usepackage{amsmath}

\usepackage{hyperref}
\usepackage{graphicx}       % Include figure files
\usepackage{dcolumn}        % Align table columns on decimal point
\usepackage{bm}             % bold math
\usepackage{amssymb}
\usepackage{amstext}
\usepackage{tensor}

\date{\today}

\newcommand{\insertplot}[5]{\begin{figure}
 \hfill\hbox to 0.05in{\vbox to #5in{\vfill
 \inputplot{#1}{#4}{#5}}\hfill}
 \hfill\vspace{-.1in}
 \caption{#2}\label{#3}
 \end{figure}}
 \newcommand{\inputplot}[3]{
 \special{ps: plotfile #1}
}
\newcounter{fig}

\newcommand{\ee}{\end{equation}}
\newcommand{\eea}{\end{eqnarray}}
\newcommand{\bea}{\begin{eqnarray}}

\newcommand{\gam}{\gamma}
\newcommand{\bet}{\beta}
\newcommand{\alp}{\alpha}

\newcommand{\sig}{\sigma}

\newcommand{\beq}{\begin{equation}}
\newcommand{\eeq}{\end{equation}}

\begin{document}

  \begin{center}

{\LARGE \bf Asymptotically flat,
 spherical, self-interacting
\\
scalar, Dirac and Proca stars
%\\
%a comparative study
}
\vspace{0.8cm}
\\
{{\bf
Carlos A. R. Herdeiro and
Eugen Radu
}
\vspace{0.3cm}
\\
$^{\ddagger }${\small Departamento de Matem\'atica da Universidade de Aveiro and } \\ {\small  Centre for Research and Development  in Mathematics and Applications (CIDMA),} \\ {\small    Campus de Santiago, 3810-183 Aveiro, Portugal}
}
\vspace{0.3cm}
\end{center}

\date{September 2020}

\begin{abstract}
We present a comparative analysis of the self-gravitating solitons  arising in
the Einstein-Klein-Gordon, Einstein-Dirac and Einstein-Proca models,
for the particular
case of static, spherically symmetric spacetimes.
Differently from the previous study \cite{Herdeiro:2017fhv},
the matter fields possess suitable self-interacting terms in the Lagrangians,
which allow for the existence of $Q$-ball--type solutions  for these models in the  flat spacetime limit.
In spite of this important difference,
our analysis shows that the high degree of universality observed in \cite{Herdeiro:2017fhv}
remains, and various spin-independent common patterns are observed.

\end{abstract}

 \tableofcontents
	
%%%%%%%%%%%%%%%%%%%%%%%%%%%%%%%%%%%%%%%%%%%
\section{Introduction and motivation}
%%%%%%%%%%%%%%%%%%%%%%%%%%%%%%%%%%%%%%%%%%%	

%%%%%%%%%%%%%%%%%%%%%%%%%%%%%%%%%%%%%%%%%%%
\subsection{General remarks}
%%%%%%%%%%%%%%%%%%%%%%%%%%%%%%%%%%%%%%%%%%%	

The (modern) idea of solitons as extended particle-like configurations can be traced back
(at least)
 to Lord Kelvin,
around one and half centuries ago,
who proposed that atoms are made of vortex knots \cite{Kelvin}.
The first
explicit example of solitons in a relativistic field theory, however, was found by
Skyrme~\cite{Skyrme:1961vq},  (almost) one hundred years later.
The latter, dubbed \textit{Skyrmions}, exist in a model with four (real) scalars
subject to a constraint.
They were proposed as a field
theory realizations of baryons; nonetheless Skyrmions
capture the basic properties
of a generic soliton.

In the context of this work,
solitons are defined as spatially localized,
 singularity-free solutions of a field theory model,
 which possess a finite
 mass and angular momentum\footnote{The fields may possess a non-trivial time-dependence, as for the solitons in this work.
However, this dependence is absent at the level of
the energy-momentum tensor.
Also, in our definition, we shall ignore the issue of stability and do not impose the solutions
possess a topological charge.
}.
As already suspected in the literature before Skyrme's seminal work\footnote{In a historical setting,
a concrete realization of
the idea of (relativistic) solitons was put forward   in 1939 by Rosen
 \cite{Rosen:1939zz}, who looked for localized configurations
 which could be used as representations of extended particles.}
many non-linear  field theories possess such solutions.
The study of solitons has
attracted a lot of attention in the last 50 years; they have
important
applications
in various contexts ranging
from   $e.g.$  the models of condensed matter physics
to high energy physics and cosmology.
Moreover, solitons are also significant for a nonperturbative quantum description of states.
Famous examples of solitons include the sphalerons in the Standard Model
\cite{Klinkhamer:1984di}
and
the magnetic monopoles \cite{tHooft:1974kcl} in various extensions thereof.

\medskip

A review of soliton physics in the non-gravitating case can be found in the textbooks
\cite{Manton:2004tk,Shnir:2018yzp};
see also the reviews
\cite{Lee:1991ax, Radu:2008pp}.
In this work,
  we shall focus on a special class of solitons in four spacetime dimensions
	that are time-dependent (at the level of the matter field)
but non-dissipative (with a time independent spacetime  geometry),
and carrying
a Noether charge $Q$ associated with
a {\it global} $U(1)$ symmetry.  We shall restrict ourselves to models with a single matter field
(except for the fermionic case)   and
  possessing  a standard kinetic term.
These solitons do not possess topological properties; but the field theory Lagrangian must have
suitable self-interactions.

Three different models are considered  here: (ungauged) field theories describing particles of spin $s=0,1/2,1$,  with the goal of providing  a comparative analysis of their  spherically symmetric solitonic solutions. 
As such, this work is an extension of that in Ref.
\cite{Herdeiro:2017fhv},  which  dealt with the same matter fields, but without self-interactions.
The latter introduce a  novel feature: self-interactions allow the existence of solitonic solutions even on a fixed flat spacetime background, $i.e.$ ignoring gravity.
As we shall see, self-interactions lead to a more intricate landscape of (gravitating)
solutions,
with some new qualitative features as compared to the picture revealed in \cite{Herdeiro:2017fhv}.

\medskip

Historically, in the non-gravitating case,  solitons in a
  complex scalar field model have been known for more than 50 years, being
	dubbed $Q-balls$   by Coleman~\cite{Coleman:1985ki}
(see also the previous work  \cite{Friedberg:1976me}). These are central  for our discussion,
sharing most of the basic features with the
higher spin generalizations.
Within a Dirac field model with a positive quartic self-interaction term,
Ivanenko \cite{Ivanenko},
Weyl \cite{Weyl},
Heisenberg
\cite{Heisenberg}
and Finkelstein et. $al.$
\cite{Finkelstein:1951zz,Finkelstein2},
have made early attempts to construct particle-like solutions.
The first rigorous
numerical study of such configurations, however,
has been done by Soler in 1976
 \cite{Soler:1970xp}.
The case of a (Abelian) vector field  has been ignored until recently;
the study of  $Q$-ball--like
solutions was initiated by
 Loginov~\cite{Loginov:2015rya}.

In all three cases, a large literature has grown based on these early studies,
which includes generalizations in various directions.
A particularly interesting case concerns the study of backreaction of these solutions on
the spacetime geometry.
This  is a legitimate question,
since the existence of solitons
 relies  on the nonlinearity of the field theory and General
Relativity (GR)
 is intrinsically highly nonlinear. Therefore, gravitational interactions could significantly alter
the flat spacetime solitons.

The results in the literature prove that all flat space
 solitons survive  when including gravity effects;
however, a more complicated picture emerges
as compared to the non-self--interacting case.
The study of gravitating generalization of the
flat spacetime $Q$-balls has started with Friedberg et al.~\cite{Friedberg:1986tq}.
The higher spin fields are less studied;
GR generalization of the self-interacting spinors were considered only recently
by {Dzhunushaliev and Folomeev
\cite{Dzhunushaliev:2018jhj},
while
the gravitating $Q$-Proca stars have been studied by Brihaye et al.
\cite{Brihaye:2017inn} and Minamitsuji~\cite{Minamitsuji:2018kof}.

\medskip

The main purpose of this work is to review various
existing results on this type of solitons,
putting them together under a consistent set of notations and conventions.
A comparative study is presented, starting with the generic framework and scaling symmetries
(Section 2).
There and afterwards,
the mathematical description of each of the
three models is made in parallel to emphasise their similarities.
The basic properties of solutions in a flat spacetime background are discussed in
Section 3, where we adapt Deser's argument \cite{Deser:1976wq} to find virial identities
which yield some insight on the existence of such configurations.
Gravity is included in Section 4; a curved spacetime generalization of
the Deser-type virial identities is also discussed.
Extending the models' framework to include GR effects allows for the possibility
to add a black hole (BH) horizon at the center of soliton, a possibility
which is indeed realized for other well-known solutions, such as Skyrmions,
 magnetic monopoles and sphalerons
(see $e.g.$ the review work \cite{Volkov:1998cc}).
However, as we shall discuss in Section 4 (and Appendix B), the situation is different for
the considered (spherically symmetric) solutions, and
there are no BH generalizations, regardelss of the presence of self-interactions.
We also discuss the
  physical interpretation of the fermionic solutions.
Concluding remarks and some open questions are presented in Section 5.

%%%%%%%%%%%%%%%%%%%%%%%%%%%%%%%%%%%%%%%%%%%
\subsection{Conventions}
%%%%%%%%%%%%%%%%%%%%%%%%%%%%%%%%%%%%%%%%%%%
Throughout, we set $c = \hbar= 1$ and adopt a metric signature $+2$.
 Greek letters $\alpha, \beta, \gamma \ldots$ are used for coordinate indices, whereas latin letters $a,b,c,\ldots$ are used  for tetrad basis indices. Symmetrization and antisymmetrization of indices is denoted with round and square brackets, $()$ and $[]$, respectively.
We use $\partial_\mu$, $\nabla_\mu$ and $\hat{D}_\mu$ to denote partial, covariant and spinor derivatives, respectively.

The flat spacetime metric in spherical coordinates reads
\begin{eqnarray}
\label{flat}
ds^2=-dt^2+dr^2+r^2(d\theta^2+\sin^2\theta d\varphi^2),
\end{eqnarray}
with $x^0=t$, $x^1=r$, $x^2=\theta$ and $x^3=\varphi$.

The conventions for scalars are those in Ref. \cite{Herdeiro:2015gia}.
In the Proca field case, we shall use the notation and conventions in
 \cite{Brito:2015pxa,Herdeiro:2016tmi}.
For fermions we shall follow the framework (including the definitions and  conventions)
in \cite{Dolan:2015eua}, as reviewed in Appendix A.

For the matter content we shall consider three different fields  $\psi$, with the specific form:
\begin{itemize}
\item
{\bf spin 0:}
\\
$\Phi$ is a complex scalar field, which is
 equivalent to a model with two real scalar fields, $\Phi^R,\Phi^I$, via  $\Phi=\Phi^R+i\Phi^I$.
 \item
{\bf spin 1/2:}
\\
$\Psi^{(A)}$ are  massive spinors, with four complex components, the index $(A)$
corresponding to the number of copies of the Lagrangian.
For a spherically symmetric configuration,
one should consider (at least) two spinors ($A=1,2$), with equal mass $\mu$.
A model with a single spinor necessarily possesses a nonzero angular momentum density and cannot be
spherically symmetric.
\item
{\bf spin 1:}
\\
$\mathcal{A}$ is a complex 4-potential, with the field strength $\mathcal{F} =d\mathcal{A}$.
Again, the model can be described in terms of two real vector fields,  $\mathcal{A}=\mathcal{A}^R+i\mathcal{A}^I$.
\end{itemize}

\medskip

Some relations get a simpler form by defining
\begin{eqnarray}
\label{psi2}
\psi^2\equiv \{ \Phi^*\Phi;~\mathcal{A}_\alpha\bar{\mathcal{A}}^\alpha ; ~  i \overline{\Psi}^{(A)}  \Psi^{(A)} \} \  ,
\end{eqnarray}
for spin $0,1/2$ and $1$, respectively,
	while we denote $\psi^4\equiv (\psi^2)^2$ and $\psi^6\equiv (\psi^2)^3$.
	
	\medskip
	The numerical construction of the  solutions reported in this work is standard.
	In our approach, we use a
Runge-Kutta ordinary differential equation solver.
 For each model,
we evaluate the initial conditions
close to the origin
at $\epsilon =  10^{-6}$ for global
tolerance $10^{-14}$, adjusting for shooting parameters\
 (which are some constants which enter the near
origin expansion of the solutions) and integrating towards $r\to \infty$.
The accuracy of the solutions was also
monitored by computing virial relations satisfied by these systems, as discussed below.
For a given set of input parameters, the solutions form a discrete
set labelled by the number of nodes, $n$, of (some of) the matter function(s).
Only fundamental states, which have the minimal number of radial nodes, are considered in this work.
	
%%%%%%%%%%%%%%%%%%%%%%%%%%%%%%%%%%%%%%%%%%%
\section{The general framework}
%%%%%%%%%%%%%%%%%%%%%%%%%%%%%%%%%%%%%%%%%%%

%%%%%%%%%%%%%%%%%%%%%%%%%%%%%%%%%%%%%%%%%%%
\subsection{The action and field equations}
%%%%%%%%%%%%%%%%%%%%%%%%%%%%%%%%%%%%%%%%%%%
We consider Einstein's gravity minimally coupled to spin$-s$
matter
fields $\psi_{(s)}=\{ \Phi;   \Psi ; \mathcal{A} \}$, with $s=0,1/2,1$. The action reads
\begin{eqnarray}
\label{action}
S=
\frac{1}{4 \pi  }\int d^4 x \sqrt{-g}
\left [
\frac{1}{4 G} R
-
L_{(s)}
\right],
\end{eqnarray}
where
\begin{eqnarray}
\label{Ltot}
L_{(s)} =L_{(s)}^{(0)}+ U_{(s)}^{\rm (int)}~;
\end{eqnarray}
the kinetic part of  the matter Lagrangians reads, respectively,
\begin{eqnarray}
\label{LS}
\nonumber
%&&{\rm scalar:}~~~L_{(0)}= - g^{\alpha \beta}\Phi_{, \, \alpha}^* \Phi_{, \, \beta} - \mu^2 \Phi^*\Phi,
&&{\rm scalar:}~~~L_{(0)}^{(0)}=   g^{\alpha \beta}\Phi_{, \, \alpha}^* \Phi_{, \, \beta},
 %+ \mu^2 \Phi^*\Phi
%-U_{int}(\Phi^*\Phi),
\\
%\end{eqnarray}
%
%\begin{eqnarray}
\label{LD}
&&{\rm Dirac:}~~~
 L_{(1/2)}^{(A)(0)} =  i
\left[
\frac{1}{2}
  \left( \{ \hat{D}  \overline{\Psi}^{(A)}  \} \Psi^{(A)} -
     \overline{\Psi}^{(A)} \hat{D}  \Psi^{(A)}
	\right)
%+   \mu \overline{\Psi}^{(A)}  \Psi^{(A)}
%+U_{int}(\overline{\Psi}^{(A)}  \Psi^{(A)})
\right],
%\end{eqnarray}
%%
%\begin{eqnarray}
\\
\label{LP}
\nonumber
 &&{\rm Proca:}~~~ L_{(1)}^{(0)}=
 \frac{1}{4}\mathcal{F}_{\alpha\beta}\bar{\mathcal{F}}^{\alpha\beta}~.
%+\frac{1}{2}\mu^2\mathcal{A}_\alpha\bar{\mathcal{A}}^\alpha~.
%-U_{int}(\mathcal{A}_\alpha\bar{\mathcal{A}}^\alpha)~.
\end{eqnarray}
$U_{(s)}^{\rm (int)}$ in (\ref{Ltot}) is a potential term,  which includes  self-interactions.
%which included the field(s) self-interaction.
In what follows we shall consider the
simplest
form of  $U_{(s)}^{\rm (int)}$
which allows for solitonic solutions in a flat spacetime background, with polynomial terms.
As such, apart from the mass term,
$U_{(s)}^{\rm (int)}$
contains quartic and sextic terms only.
Then all three cases can be treated
in a unitary way by defining
\begin{eqnarray}
\label{U}
U_{(s)}^{\rm (int)}= \mathcal{M} \psi^2-\lambda \psi^4+\nu \psi^6 \equiv U   \ ,
\end{eqnarray}
where we  have  denoted
\begin{eqnarray}
\label{M}
 \mathcal{M} \equiv \left\{ \mu^2;~\mu; ~~\frac{1}{2}\mu^2  \right\} \ ,
\end{eqnarray}
for spin $0,1/2$ and $1$, respectively.
In all cases, $\mu$ corresponds to the mass of the elementary quanta of the field(s).
Also, $\lambda,~\nu$
are input parameters of the theory,
which are not fixed \textit{a priori}.

We shall also denote
\begin{equation}
\label{exprU}
\dot{U}\equiv \frac{\partial U}{\partial \psi^2}= \mathcal{M}-2\lambda \psi^2+3 \nu \psi^4,~~ \qquad
\ddot{U}\equiv\frac{\partial  \dot{U}}{\partial \psi^2}=-2\lambda +6 \nu \psi^2.
\end{equation}

%%%%%%%%%%%%%%%%%%%%%%%%%%%%%%%%%%%%%%%%%%%
%\subsection{The field equations}
%%%%%%%%%%%%%%%%%%%%%%%%%%%%%%%%%%%%%%%%%%%

\medskip

Extremizing the action (\ref{action}) leads to the Einstein field equations
\begin{eqnarray}
R_{\alpha \beta}-\frac{1}{2}Rg_{\alpha \beta}=2 G~T_{\alpha \beta} \ ,
\end{eqnarray}
where the energy-momentum  tensor reads, respectively,
\begin{eqnarray}
\label{TS}
\nonumber~
&&{\rm scalar:}~~
T_{\alpha \beta}=
 \Phi_{ , \alpha}^*\Phi_{,\beta}
+\Phi_{,\beta}^*\Phi_{,\alpha}
-g_{\alpha \beta} L_{(0)}  ,
%\end{eqnarray}
%
%
%\begin{eqnarray}
\\
&&{\rm Dirac:}~
T_{\alpha \beta}=\sum_A T_{\alpha \beta}^{(A)},~~{\rm with}~~T_{\alpha \beta}^{(A)} =-\frac{i}{2}
\left[
    \overline{\Psi}^{(A)} \gamma_{(\alpha} \hat{D}_{\beta)} \Psi^{(A)}
-  \left\{ \hat{D}_{(\alpha} \overline{\Psi}^{(A)} \right\} \, \gamma_{\beta)} \Psi^{(A)}
- g_{\alpha \beta} L_{(1/2)}
\right],~~~~{~~~~}  \nonumber
\label{TD}
\\
&&{\rm Proca:}~
T_{\alpha\beta}=\frac{1}{2}
( \mathcal{F}_{\alpha \sigma }\bar{\mathcal{F}}_{\beta \gamma}
+\bar{\mathcal{F}}_{\alpha \sigma } \mathcal{F}_{\beta \gamma}
)g^{\sigma \gamma}
+\dot{U}
(\mathcal{A}_{\alpha}\bar{\mathcal{A}}_{\beta}
 +\bar{\mathcal{A}}_{\alpha}\mathcal{A}_{\beta})
-g_{\alpha\beta}L_{(1)}  ,~{~~}
\label{TP}
\end{eqnarray}
and to the matter field equations,
\begin{eqnarray}
\label{ES}
\nonumber~
&&{\rm scalar:}~~(\nabla^2 -\dot{U} )\Phi =0 ,
\\
&&{\rm Dirac:}~~
    (\hat{D}   - \dot{U}  )\Psi^{(A)}   = 0 ,
\\
\nonumber~
&&{\rm Proca:}~~
\frac{1}{2}\nabla_\alpha\mathcal{F}^{\alpha\beta}- \dot{U}  \mathcal{A}^\beta=0.
\end{eqnarray}
In the Proca case,
taking the 4-divergence of the field eqs. (\ref{ES}) results in a generalization of the Lorenz condition, which is a dynamical requirement, rather than a gauge choice. This condition takes a particularly simple form in the Ricci-flat case, with
\begin{equation}
\label{lorentz}
(\nabla_\alpha\mathcal{A}^\alpha ) \dot{U} +\ddot{U} \nabla_\alpha (\bar{\mathcal{A}}_{\beta}\mathcal{A}^{\beta})= 0 \ .
\end{equation}

%%%%%%%%%%%%%%%%%%%%%%%%%%%%%%%%%%%%%%%%%%%
%\subsection{The current and Noether charge}
%%%%%%%%%%%%%%%%%%%%%%%%%%%%%%%%%%%%%%%%%%%

In all cases, the action of the matter fields $\psi$ possesses a
  global $U(1)$ invariance, under the transformation $\psi \rightarrow e^{i\alpha}\psi $, with $\alpha$ constant.
This	implies the existence of a conserved 4-current,
which reads
\begin{eqnarray}
\label{jS}
\nonumber~
&&{\rm scalar:}~~
j^\alpha=-i (\Phi^* \partial^\alpha \Phi-\Phi \partial^\alpha \Phi^*),
\\
&&{\rm Dirac:}~~j^\alpha=\bar \Psi \gamma^\alpha \Psi ,
\\
\nonumber~
&&{\rm Proca:}~~j^\alpha=
\frac{i}{2}\left[\bar{\mathcal{F}}^{\alpha\beta}\mathcal{A}_\beta-\mathcal{F}^{\alpha \beta}\bar{\mathcal{A}}_\beta\right].
\end{eqnarray}
This current is conserved via the field equations,
\begin{eqnarray}
j^\alpha_{\ ;\alpha}=0.
\end{eqnarray}
It follows that integrating the timelike component of this 4-current in a spacelike slice $\Sigma$ yields a conserved quantity --
 the \textit{Noether charge}:
\begin{eqnarray}
\label{Q}
Q=\int_{\Sigma}~j^t \ .
\end{eqnarray}

%%%%%%%%%%%%%%%%%%%%%%%%%%%%%%%%%%%%%%%%%%%
\subsection{The Ans\"atze and explicit equations}
%%%%%%%%%%%%%%%%%%%%%%%%%%%%%%%%%%%%%%%%%%%
%%%%%%%%%%%%%%%%%%%%%%%%%%%%%%%%%%%%%%%%%%%
\subsubsection{The metric and matter fields}
%%%%%%%%%%%%%%%%%%%%%%%%%%%%%%%%%%%%%%%%%%%
The
spherically symmetric configurations are most conveniently studied in Schwarzschild-like coordinates,
within the following metric Ansatz:
\begin{eqnarray}
\label{metric}
 ds^2=-N(r)\sigma^2(r) dt^2+\frac{dr^2}{N(r)}+r^2 (d\theta^2+\sin^2\theta d\varphi^2)\ ,
~~~{\rm with}~~N(r)\equiv 1-\frac{2m(r)}{r}.
\end{eqnarray}

The matter field Ansatz which is compatible with a spherically symmetric geometry reads:
\begin{eqnarray}
\label{S}
&&{\rm scalar:}~~
\Phi=\phi(r)e^{-iw t},
\end{eqnarray}
introducing a single real function $\phi(r)$;
\begin{eqnarray}
&&{\rm Proca:}~~
\mathcal{A}=\left[F(r)dt+iH(r)dr\right] e^{-iwt},
\end{eqnarray}
introducing two real potentials $F(r)$ and $H(r)$.

The case of a Dirac field is more involved.
For a spherically symmetric configurations we have to consider two Dirac fields, $A=1,2$, with
\begin{eqnarray}
\label{nn1}
&&{\rm Dirac:}~~
\Psi^{(A)} = \mathrm e^{-\mathrm i w t}
{\cal R}^{(A)}(r)
\otimes \Theta^{(A)} (\theta,\varphi) \, ,
\end{eqnarray}
where
\begin{eqnarray}
\label{nn2}
{\cal R}^{(1)} =-i{\cal R}^{(2)} =
 \bigg[
                          \begin{array}{c}
z (r)
\\
-i \bar z(r)
                          \end{array}
 \bigg] ~,
~~~
\nonumber
\Theta^{(1)} = \bigg[
                          \begin{array}{c}
-\kappa \sin \frac{\theta}{2}
\\
\cos \frac{\theta}{2}
                          \end{array}
 \bigg] e^{i\frac{\varphi}{2}},~~
\Theta^{(2)} = \bigg[
                          \begin{array}{c}
 \kappa \cos \frac{\theta}{2}
\\
\sin \frac{\theta}{2}
                          \end{array}
 \bigg] e^{-i\frac{\varphi}{2}},~~
\end{eqnarray}
with $\kappa=\pm 1$ and $z(r)$ a complex function.
In what follows, we shall consider the case $\kappa= 1$ only
(note that qualitatively similar solutions with $\kappa= -1$
exist as well).
Also, in the above Ansatz it is convenient to define
\begin{eqnarray}
\label{z}
z(r)=   e^{i\pi/4} f(r)-e^{-i\pi/4} g(r)~,
\end{eqnarray}
where $f(r)$ and $g(r)$ are two real functions, a choice which simplifies the equations.

The reason we choose two independent  $s=1/2$ fields, with the Ansatz (\ref{nn1}), (\ref{nn2}),
 is the following.
For both spinors, the $individual$ energy-momentum tensors are not spherically symmetric\footnote{This feature
is present regardless of the  self-interaction potential in the Lagrangian.}, since
$T_\varphi^{t(A)}\neq 0$
(while the other nonzero components
$T_{r}^{r(A)}$,
$T_{\theta}^{\theta(A)}$=
$T_{\varphi}^{\varphi(A)}$
and
$T_{t}^{t(A)}$
depend on $r$ only).
However, $T_\varphi^{t(1)}+T_\varphi^{t(2)}=0$, such that the full configuration is spherically symmetric,
being compatible with the line element (\ref{metric}).

Within this framework, the explicit expressions of $\psi^2$,
as defined in
(\ref{psi2}),
are
\begin{eqnarray}
\label{psi2z}
 \psi^2= \left\{ \phi^2,~~4(g^2-f^2),~~H^2N-\frac{F^2}{N\sigma^2}\right\}
\end{eqnarray}
for spin $0,1/2$ and $1$, respectively.
In all case, $w>0$ is the frequency of the matter field.

%%%%%%%%%%%%%%%%%%%%%%%%%%%%%%%%%%%%%%%%%%%
\subsubsection{The explicit equations }
%%%%%%%%%%%%%%%%%%%%%%%%%%%%%%%%%%%%%%%%%%%
The equations for the mass function $m(r)$ read, respectively,
\begin{eqnarray}
\label{m-S}
\nonumber~
&&{\rm scalar:}~~
 m'=  G r^2
\left(
N \phi'^2+\frac{w^2 \phi^2}{N\sigma^2}
+U
\right),
\\
\label{m-D}
&&{\rm Dirac:}~~
% m'=8 G r^2
%\left(
%\frac{w(f^2+g^2)}{\sqrt{\sigma}N}
%+\frac{1}{4}(\lambda \psi^4-2\nu \psi^6 )
%\right ),
%
 m'=2 G r^2
\left(
4\sqrt{N}(gf'-fg')+\frac{8fg}{r}+ U
\right ),
\\
\nonumber~
\label{m-P}
&&{\rm Proca:}~~
m'= G r^2
\left[
\frac{(F'-wH)^2}{ 2\sigma^2}
+ (\mu^2-6\lambda {\cal A}^2+10\nu  {\cal A}^4)\frac{F^2}{2N\sigma^2}
+\frac{ U}{{\cal A}^2}N G^2
\right].
\end{eqnarray}

The equations for the metric function $\sigma(r)$ read, respectively,
\begin{eqnarray}
\label{s-S}
\nonumber~
&&{\rm scalar:}~~
\frac{\sigma'}{\sigma}
=2 G r
\left(
\phi'^2+\frac{w^2\phi^2}{N^2\sigma^2}
\right),
\\
&&{\rm Dirac:}~~
\frac{\sigma'}{\sigma}= 8 G\frac{r}{\sqrt{N}}
\left(
gf'-fg'+\frac{w(f^2+g^2)}{N\sigma}
\right),
\\
\nonumber~
&&{\rm Proca:}~~
\frac{\sigma'}{\sigma}=
\frac{2 G r}{N}
%\frac{dU}{d{\cal A}^2}
\left(
H^2 N +\frac{F^2}{N\sigma^2}
\right)
\dot U.
\end{eqnarray}
The equations for the matter fields are
\begin{eqnarray}
\label{e-S}
\nonumber~
&&{\rm scalar:}~~
\phi''+\left(\frac{2}{r}+\frac{N'}{N}+\frac{\sigma'}{\sigma}\right)\phi'
+
%(\frac{w^2}{N\sigma^2}-\mu^2)\frac{\phi}{N}+(2\lambda -3\nu \phi^2)\frac{\phi^3}{N} =0,
\frac{w^2}{N^2\sigma^2}\phi -\dot U \frac{\phi}{N}=0,
\\
\nonumber~
&&{\rm Dirac-f:}~~
f'+
\left(
\frac{N'}{4N}+\frac{\sigma'}{2\sigma}+\frac{1}{r\sqrt{N}}+\frac{1}{r}
\right)f
%(
%\frac{\mu}{\sqrt{N}}-\frac{w}{ N\sigma}
%)g
%+\frac{8  g(f^2-g^2)}{\sqrt{N}}\left(\lambda+6\nu (f^2-g^2) \right)=0,~~~{~~~}
-\frac{w g}{ N\sigma}
+\frac{g}{\sqrt{N}}\dot U=0,
%+\frac{8  g(f^2-g^2)}{\sqrt{N}}\left(\lambda+6\nu (f^2-g^2) \right)=0,~~~{~~~}
\\
&&{\rm Dirac-g:}~~
g'+
\left(
\frac{N'}{4N}+\frac{\sigma'}{2\sigma}-\frac{1}{r\sqrt{N}}+\frac{1}{r}
\right)f
%(
%\frac{\mu}{\sqrt{N}}+\frac{w}{ N\sigma}
%)g
%+\frac{8 f(f^2-g^2)}{\sqrt{N}}\left(\lambda+6\nu (f^2-g^2) \right)=0,~~~{~~~}
+\frac{w f}{ N\sigma}
+\frac{f}{\sqrt{N}}\dot U=0,
\\
\nonumber~
&&{\rm Proca-F:}~~
  \qquad
	%F'- wH+\frac{2 N \sigma^2 H } {w}\frac{dU}{d{\cal A}^2}=0~,
	F'- wH+\frac{2 N \sigma^2 H } {w}\dot U=0~,
\\
\nonumber~
&&{\rm Proca-H:}~~
 \frac{d}{dr}\left\{\frac{r^2[w H-F' ]}{\sigma}\right\}+\frac{ 2r^2F}{N \sigma }
%\frac{dU}{d{\cal A}^2}=0~.
\dot U =0~.
\end{eqnarray}
An  additional supplementary constraint is also present, which is a 2nd order equation
for the metric functions $m(r)$ and $\sigma(r)$, with first order derivatives of matter fields.
However, this equation is a differential consequence of the above field equations; it is used to check the
accuracy of the numerical results.

%%%%%%%%%%%%%%%%%%%%%%%%%%%%%%%%%%%%%%%%%%%%%%%%%%
%\subsection{The effective action(s) }
%%%%%%%%%%%%%%%%%%%%%%%%%%%%%%%%%%%%%%%%%%%%%%%%%%
 \medskip

Let us remark that the above equations can be derived from the following
effective action
\begin{eqnarray}
\label{eff-action}
S_{\rm eff}=\int dr {\cal L}_{\rm eff} \ , \qquad ~~~ {\rm with}~~~
{\cal L}_{\rm eff}=
\frac{1}{ G}\sigma m'- {\cal L}_{(s)} \ ,
\end{eqnarray}
where
\begin{eqnarray}
\label{Leff0}
\nonumber~
&&
 {\cal L}_{(0)}=r^2\sigma
\left(
N \phi'^2-\frac{w^2 \phi^2}{N \sigma^2}+U
\right),
\\
&&
\label{LeffD}
 {\cal L}_{(1/2)}=8 r^2\sigma
\left(
 \sqrt{N}(gf'-fg')
-\frac{w}{\sigma \sqrt{N}}(f^2+g^2)
+\frac{2 fg}{r}
+\frac{1}{4}U
\right),
\\
\label{Leff1}
\nonumber~
&&
 {\cal L}_{(1)}=r^2\sigma
\left(
-\frac{(F'-w H)^2}{2\sigma^2}
+U
\right).
\end{eqnarray}

Given the above framework,
the energy density measured by a static observer, $\rho=-T_t^t$, is
\begin{eqnarray}
\label{rho-S}
\nonumber~
&&{\rm scalar:}~~
\rho=N\phi'^2+ \frac{w^2}{N\sigma^2} \phi^2+U~,
\\
&&{\rm Dirac:}~~
\rho=
8\left(
(gf'-fg')\sqrt{N}+\frac{2fg}{r}+\frac{1}{4}U
\right)~,
\\
\nonumber~
&&{\rm Proca:}~~
\rho= \frac{(F'-wH)^2}{2\sigma^2}
+ \frac{ U}{{\cal A}^2}NH^2
+\frac{F^2}{N\sigma^2}
\left(
\frac{1}{2}\mu^2-3\lambda {\cal A}^2+5\nu {\cal A}^4
\right)
.
\end{eqnarray}
The mass of the flat space solutions is computed
as the integral of $\rho$,
\begin{eqnarray}
M=\int dr r^2 \rho,
\end{eqnarray}
while in the self-gravitating case
it can be read off from the asymptotic behaviour of the $g_{tt}$
metric potential
\begin{eqnarray}
g_{tt}=-N\sigma^2=-1+\frac{2MG}{r}+\dots
\end{eqnarray}
As for the Noether charge, it reads
\begin{eqnarray}
\label{Q-S}
\nonumber~
&&{\rm scalar:}~~
%Q=8\pi w \int_0^\infty dr \,r^2  \frac{\phi^2}{N\sigma},
Q=2 w \int_0^\infty dr \,r^2  \frac{\phi^2}{N\sigma},
\\
\label{Q-D}
&&{\rm Dirac:}~~
%Q= 16\pi \int_0^\infty dr \,r^2  \frac{(f^2+g^2)}{\sqrt{N}},
Q=4 \int_0^\infty dr \,r^2  \frac{(f^2+g^2)}{\sqrt{N}},
\\
\nonumber~
\label{Q-P}
&&{\rm Proca:}~~
%Q= 8\pi \int^{\infty}_0 dr\, r^2 \frac{(wG-F')G}{\sigma}~.
Q= 2 \int^{\infty}_0 dr\, r^2 \frac{(wH-F')N}{\sigma}~.
\end{eqnarray}

%%%%%%%%%%%%%%%%%%%%%%%%%%%%%%%%%%%%%%%%%%%
\subsection{Units and scaling symmetries}
%%%%%%%%%%%%%%%%%%%%%%%%%%%%%%%%%%%%%%%%%%%

%%%%%%%%%%%%%%%%%%%%%%%%%%%%%%%%%%%%%%%%%%%
%\subsection{Units }
%%%%%%%%%%%%%%%%%%%%%%%%%%%%%%%%%%%%%%%%%%%

The matter fields in action (\ref{action}) have the following dimensions:
(with $L=lenght$):
\begin{eqnarray}
\nonumber~
&&{\rm scalar:}~~
[\Phi]=\frac{1}{L},
\\
&&{\rm Dirac:}~~[\Psi]=\frac{1}{L^{3/2}},~~
\\
\nonumber~
&&{\rm Proca:}~~
[\mathcal{A}_a] =\frac{1}{L}.
\end{eqnarray}
Also,
 the coupling constants which enter the potential are (generically) dimension-full, with:
\begin{eqnarray}
\nonumber~
&&{\rm scalar:}~~
[\mu]=\frac{1}{L},~~[\lambda]=L^0,~~[\nu]=L^2,
\\
&&{\rm Dirac:}~~[\mu]=\frac{1}{L},~~[\lambda]=L^2,~~[\nu]=L^5,
\\
\nonumber~
&&{\rm Proca:}~~
[\mu]=\frac{1}{L},~~[\lambda]=L^0,~~[\nu]=L^2.
\end{eqnarray}

%%%%%%%%%%%%%%%%%%%%%%%%%%%%%%%%%%%%%%%%%%%%%%%%%%%%%%
%\subsection{Scaling symmetries }
%%%%%%%%%%%%%%%%%%%%%%%%%%%%%%%%%%%%%%%%%%%%%%%%%%%%%%

Turning now to scaling symmetries,
we notice the existence of $three$
different transformations which,
in all cases, leave invariant the equations of motion
(in the relations below, the functions or constants which are not mentioned explicitly remain invariant).

The first one (s0) is very simple
\begin{eqnarray}
\label{s0}
(s0):~~~\sigma \to c \sigma,~~w\to c w,
\end{eqnarray}
with $c$ some arbitrary positive constant.
%while the other functions and input constants  are not changed.
However,
this invariance
%(which is important in the numerics)
is fixed when imposing the asymptotic flatness condition
$\sigma \to 1$
as $r\to \infty$.

More importantly,
the equations of motion are invariant under a scaling of the radial coordinate
together with other functions  and parameters of the model.
 %with $c$ some arbitrary positive constant
In the scalar and Proca cases, this transformation reads:
\begin{eqnarray}
\label{s1}
(s1):~~~{\rm scalar~and~Proca:}~~
  r= c \bar r,~~w=\frac{1}{c}\bar w,~~
 ~\mu=\frac{1}{c}\bar \mu, ~~\lambda=\frac{\bar \lambda }{c^2},~~\nu=\frac{\bar \nu }{c^4},~~m = c \bar m.
\end{eqnarray}
%all other functions remaining invariant.
%, except in the spin $1/2$ case, where we have in addition to (\ref{s1})
%while the scalar field $\phi(r)$ and the metric function $\sigma(r)$ remain invariant.
In the Dirac case, the corresponding symmetry is more complicated, with
\begin{eqnarray}
\label{s1-dirac}
&&(s1):~~~{\rm Dirac:}~~
r= c \bar r,~~w=\frac{1}{c}\bar w,~ \mu=\frac{1}{c}\bar \mu,~~\nu=c \bar \nu;~~~
m = c \bar m  ,~~f=\frac{\bar f}{\sqrt{c}},~~g=\frac{\bar g}{\sqrt{c}}.
\end{eqnarray}
In all cases the product $m \mu$ and the ratio $w/\mu$ are left invariant by the symmetry $(s1)$.
Under this transformation, the global quantities behave as
\begin{eqnarray}
\label{s1-MQ}
 M= c \bar M,~~Q=c^2 \bar Q.
\end{eqnarray}

The  symmetry $(s1)$ is usually employed to work in units of length set by the field mass,
\begin{eqnarray}
\bar \mu=1, ~~i.e.~~c=\frac{1}{\mu}.
\end{eqnarray}

Finally, the equations are also  invariant under a suitable scaling of the
matter field(s) functions together with some coupling constants
(while the radial coordinate or the field frequency are not affected):
\begin{eqnarray}
\label{s21}
&&(s2):~~
\lambda =\frac{\bar \lambda }{c^2},~~\nu =\frac{\bar \nu }{c^2},~~G =\frac{\bar G }{c^2} \ ,
\end{eqnarray}
together with
\begin{eqnarray}
\label{s22}
&&(s2):~~~{\rm scalar:}~~
\phi =\bar \phi c, ~~~{\rm Dirac:}~~f =\bar f c,~~g =\bar g c,~~~{\rm Proca:}~~F =\bar F c,~~H =\bar H c,
\end{eqnarray}
while
the global quantities transform as
\begin{eqnarray}
\label{s2-MQ}
 M= c^2 \bar M,~~Q=c^2 \bar Q.
\end{eqnarray}

\medskip

The symmetries  $(s1)$,  $(s2)$ are used in practice to simplify the
numerical study of the solitons.
First,
the  symmetry  $(s2)$
can be used to set $\bar G=1$; $i.e.$ essentially one absorbs Newton's constant in the expression of the matter field(s).
This is the usual  approach for the models without self-interaction,
see
$e.g.$ the discussion
 in \cite{Herdeiro:2017fhv}.

However, the probe limit of the solutions
becomes less transparent for this choice.
An alternative route, employed in this work,
is to use (\ref{s21}), (\ref{s22})
the set the coefficient of the quartic term to  unity,
\begin{eqnarray}
\bar \lambda=1, ~~i.e.~~c=\frac{1}{\sqrt{\lambda}}.
\end{eqnarray}
It follows that two mass scales naturally emerge, one set by gravity $M_{\rm Pl}=1/\sqrt{G}$
and the
other one,
$M_0$,
set by the field(s) coupling constants, with
\begin{eqnarray}
{\rm scalar~ and ~Proca:}~~ M_0=\frac{\mu}{\sqrt{\lambda}};~~~~
{\rm Dirac:}~~ M_0=\frac{1}{\sqrt{\lambda}}.
%{\rm Proca:}~~ M_0=\frac{\mu}{\sqrt{\lambda}}.
\end{eqnarray}
The ratio of these fundamental mass scales defines the dimensionless coupling constant
\begin{eqnarray}
\alpha=\frac{M_0}{M_{\rm Pl}},
\end{eqnarray}
which is relevant in the physics of the solutions.

Another dimensionless input parameter
is the  scaled constant  for the sextic term in the potential $U$, with
 \begin{eqnarray}
{\rm scalar~ and~ Proca:}~~\beta=\frac{\nu \mu^2}{\lambda^2};~~~~
{\rm Dirac:}~~ \beta=\frac{\nu \mu}{\lambda^2}.
\end{eqnarray}
As such, the mass and charge of the non-gravitating solutions
is given in units set by $\mu,\lambda$,
while in the gravitating case,
to make contact with the previous  results
without self-interaction,
 we use units set by $G$ and $\mu$.

\textit{A  priori}, the range of $\alpha$ is unbounded,
$0\leqslant \alpha< \infty$.
The limit $\alpha \to 0$ corresponds to $G\to 0$, $i.e.$ the probe limit --
one solves the matter field(s) equations on
a fixed geometry, which should be a solution of the vacuum Einstein equations.
The limit $\alpha \to \infty$ corresponds to $\lambda \to 0$, $i.e.$
no quartic or sextic term (when using  $\beta$,  if $\nu$ is finite) in the action.
Thus, solutions of the Einstein-matter field
equations without self-interaction are approached in the second limit.

This choice of units with $\mu=\lambda=1$ (after employing the above scaling symmetries)
has the advantage to
greatly simplify the numerical study of the solutions.
For example,
the Einstein equations
read
\begin{eqnarray}
R_{\alpha \beta}-\frac{1}{2}g_{\alpha \beta}=2\alpha^2 T_{\alpha \beta},
\end{eqnarray}
the only input parameters being
\begin{eqnarray}
\{ \alpha,~~ \beta ~~{\rm and}~~w\}~,
%\bar w=\frac{w}{\mu}~\}.
\end{eqnarray}
with $w$ the {\it scaled} frequency.
Also, the scaled scalar potential reads
\begin{eqnarray}
\label{scaledU}
U=\psi^2-\psi^4+\beta \psi^6,
\end{eqnarray}
For $\psi^2>0$,
  $U$ is strictly positive for  $\beta>1/4$.
The case $\beta=1/4$  is special, 
	since the potential becomes 
	$U=\psi^2(1-\psi^2/2)^2$, 
	and thus possesses
three degenerate minima, at $\psi=\{0, \pm \sqrt{2} \}$.
A discussion of these aspects 
(for a spin-zero field)
can be  found  in Ref. \cite{Brihaye:2015veu}.

%%%%%%%%%%%%%%%%%%%%%%%%%%%%%%%%%%%%%%%%%%%%%%%%%%
\section{The probe limit: flat spacetime solutions}
%%%%%%%%%%%%%%%%%%%%%%%%%%%%%%%%%%%%%%%%%%%%%%%%%%

%%%%%%%%%%%%%%%%%%%%%%%%%%%%%%%%%%%%%%%%%%%%%%%%%%%%%%%%%%%%%%%%%%%%%%%%%%%%%%
\subsection{Deser's argument and virial-type identities}
%%%%%%%%%%%%%%%%%%%%%%%%%%%%%%%%%%%%%%%%%%%%%%%%%%%%%%%%%%%%%%%%%%%%%%%%%%%%%%

Before performing a numerical study of the solutions, it is useful to derive virial-type identities.
%which help to better understand the 'raison d'etre' for the solutions.
%
For a flat spacetime background,
we can adapt a simple argument given long  ago by Deser \cite{Deser:1976wq}
(used therein to rule out the existence of finite energy time-independent solutions in Yang-Mills theory)
to obtain virial-type identities and a simple relation between the mass of solutions
and the trace of the energy-momentum tensor\footnote{The same virial identities are found by adapting Derrick's scaling
argument \cite{Derrick:1964ww}. } -  see also~\cite{Gibbons:1990um}.

Working in Cartesian coordinates $x^a$ ($a=1,2,3$),
assume the existence of a stationary soliton in some field theory model.
Following \cite{Deser:1976wq}, consider the following (trivial) identity\footnote{Observe
that Deser's argument cannot be extended to a curved spacetime background.}
\begin{eqnarray}
\label{Deser1}
\frac{\partial }{\partial x^a}\left(x^b T_b^a \right)=T_a^a+x^b \frac{\partial   T_b^a }{\partial x^a}~,
\end{eqnarray}
together with its volume integral. The left hand side vanishes from regularity and finite energy requirements\footnote{Note that in the spin-1/2 case, one considers the total energy-momentum tensor.}.
%\begin{eqnarray}
%\label{relationDeser0}
%\int d^3 x \frac{\partial (x^b T_b^a)}{\partial x^a}=0
%\end{eqnarray}
The second term in (\ref{Deser1})
vanishes from energy-momentum conservation (plus staticity)
and thus we are left with the virial-type identity
\begin{eqnarray}
\label{relationDeser1}
\int d^3 x~T_a^a=0.
\end{eqnarray}
It follows that the total mass-energy of a static soliton in $d=3+1$ dimensions
is determined by the trace of the energy-momentum tensor
\begin{eqnarray}
\label{relationDeser2}
M=-\int d^3 x ~T_t^t=-\int d^3 x ~T_\mu^\mu~.
\end{eqnarray}

When applied to the specific Ans\"atze in this work,  (\ref{relationDeser1})
results in the following expressions:
\begin{itemize}
\item
scalar field:
\end{itemize}
\begin{eqnarray}
\label{virial-scalar-flat}
 \int_0^\infty dr~ r^2
\left[
\frac{1}{3}\phi'^2+ (\mu^2 -w^2)\phi^2+  \nu \phi^6
 \right ]=
\lambda
\int_0^\infty dr~r^2
 \phi^4.
\end{eqnarray}
This relation can be simplified by using the equation for $\phi$ in a simpler form
\begin{eqnarray}
\label{virial-scalar-flat2}
 \int_0^\infty dr~ r^2
(\mu^2-w^2)\phi^2=
\frac{\lambda}{2}
\int_0^\infty dr~r^2
 \phi^4.
\end{eqnarray}
This clearly shows that {\it the (flat space) Q-ball solutions are supported by the quartic self-interacting term}, with $\lambda>0$
($i.e.$ the sextic term is not relevant at this level).
\begin{itemize}
\item
Dirac field:
\end{itemize}
\begin{eqnarray}
\label{virial-Dirac-flat}
 %\int_0^\infty dr r^2
%(
%\mu(f^2-g^2)+2 \lambda (f^2-g^2)^2
%)=w  \int_0^\infty dr r^2 (f^2+g^2)
 \int_0^\infty dr~r^2
 \left(
 gf'-fg'+\frac{2fg}{r}+\frac{3}{8}U-\frac{3}{2}w(f^2+g^2)
 \right )= 0
\end{eqnarray}
which can also be written, via field equations, in the alternative form
\begin{eqnarray}
\label{virial-Dirac-flat2}
 \int_0^\infty dr~r^2
 \left(
\mu \psi^2+\lambda \psi^4
 \right)
=
   \int_0^\infty dr~r^2
\left(4w(f^2+g^2)
+3\nu \psi^6
\right).
\end{eqnarray}
One concludes that,
for
$\nu=0$
and $\lambda>0$,
 the solutions are supported by the harmonic time dependence.
However,
the above relation does not clarify the role played by the nonlinear quartic term for the existence of solitons.

 \begin{itemize}
\item
Proca field:
\end{itemize}
\begin{eqnarray}
\label{virial-Proca-flat}
 \int_0^\infty dr~r^2
\left (
 \frac{1}{2}(F'-wG)^2-3U+2G^2\frac{dU}{d{\cal A}^2}
\right )= 0\ .
\end{eqnarray}
After eliminating the kinetic term $(F'-wG)^2$, via field equations, the above relation takes the
suggestive form:
\begin{eqnarray}
\nonumber
&&
 \int_0^\infty dr~r^2
\left[
\mu^2(3F^2+(\frac{\mu^2}{w^2}-1)G^2
+\frac{16\lambda^2 G^2}{w^2} {\cal A}^4
+\frac{6 \nu }{w^2}{\cal A}^4
\left(
w^2(F^2+G^2)+2G^2(\mu^2+3\nu {\cal A}^2
\right)
\right ]=
\\
\label{virial-Proca2}
&&{~~~~~~~~~~}
2\lambda \int_0^\infty dr~r^2
{\cal A}^2 \left (
 3F^2+G^2+ \frac{4 G^2}{w^2}( \mu^2 +6\nu  {\cal A}^2)
\right).
\end{eqnarray}
From the bound state condition $\mu^2\geqslant w^2$,
it is clear that the existence of finite mass solutions requires a quartic term, $\lambda \neq 0$ ($\nu$ being irelevant).
But since ${\cal A}^2$ may take both positive or negative values, one cannot use the above relation to
predict the sign of $\lambda$.

It is also interesting to note that the mass of the flat space Proca solitons takes the simple form
\begin{eqnarray}
\label{M-Proca-flat}
M=
 \int_0^\infty dr~r^2
\left(
 \mu^2 {\cal A}^2-2\nu {\cal A}^6
\right).
\end{eqnarray}
%For $\nu =0$, this expression clearly shows that ${\cal A}^2$ has no definite sign.

%%%%%%%%%%%%%%%%%%%%%%%%%%%%%%%%%%%%%%%%%%%
\subsection{Numerical results}
%%%%%%%%%%%%%%%%%%%%%%%%%%%%%%%%%%%%%%%%%%%

%%%%%%%%%%%%%%%%%%%%%%%%%%%%%%%%%%%%%%%%%%%
\subsubsection{General remarks}
%%%%%%%%%%%%%%%%%%%%%%%%%%%%%%%%%%%%%%%%%%%

The corresponding equations are found by taking
\begin{eqnarray}
\label{flat-limit}
N=\sigma=1 \ ,
\end{eqnarray}
in the corresponding general equations in Section 2,
and we shall not display them here.
Moreover, the boundary conditions satisfied by the functions
$\psi$
at $r=0,\infty$
are similar to those in the gravitating case, as given in the next Section.

\medskip
The case of a scalar field  is special for a flat spacetime metric.
The frequency parameter $w$ is not relevant, since
 $w^2$ acts as an effective $tachyonic$ contribution to the  mass term,
and thus it can be absorbed into $\mu^2$,
  by defining $ \mu^2-w^2\to \mu^2$.
	After this redefinition,
the scalar field is static,
$\Phi=\phi(r)$ and thus the Noether charge vanishes.
Therefore all $Q$-ball solutions in a flat spacetime background can be interpreted as $static$
scalar solitons,
in a model with a shifted scalar field mass term \cite{Kleihaus:2013tba}, for a new potential
\begin{eqnarray}
\label{new-pot}
U=U_{(w)}-w^2 \phi^2.
\end{eqnarray}
Note that although $\phi$ formally satisfies the same equation as before, the energy-momentum tensor and
the total mass of these solutions are different.
Also, the virial identities imply that
the redefined potential $U$ is necessarily negative for
some range of $\phi$, which is realised by the solutions, $U<0$.
Moreover, one can prove the following relation 
\cite{Kleihaus:2013tba}
\begin{eqnarray}
M(w=0)=M{(w)}-w Q,
\end{eqnarray}
which relates the mass of a static solution ($w=0$ and the redefined potential (\ref{new-pot}))
to the mass and Noether charge of solutions with a given $w$
(the other parameters in the potential are kept fixed).

\medskip
No similar relation seems to exist for higher spin fields.
However, it is interesting to notice the existence in this case of a curious static, purely electric solution,
$i.e.$ with $H=0$, $w=0$.
The electric potential $F(r)$ satisfies essentially the same equation
as a scalar $Q$-ball,
\begin{eqnarray}
F''+\frac{2F'}{r}-(\mu^2 +4\lambda F^2+6\nu F^4)F=0 \ .
\end{eqnarray}
The existence of solutions with proper asymptotics requires $\lambda<0$, being constructed within the same scheme as in the generic case.

To the best of our knowledge, this case has not been discussed in the literature.
However, they possess some unphysical features.
For example, by using the virial identity
(\ref{M-Proca-flat})
together with the above equations one can prove
that the total mass of this configuration is negative\footnote{Moreover, we have verified that $M$
remains negative when
including gravity effects.}
\begin{eqnarray}
M=-\frac{1}{3}\int_0^\infty dr r^2 F'^2<0.
\end{eqnarray}

 \medskip

In all cases,
the sextic self-interaction term is {\bf not} necessary for the existence of solutions.
Since the $\beta=0$ case  has some special properties, we shall discuss it separately at the end of this section.

%%%%%%%%%%%%%%%%%%%%%%%%%%%%%%%%%%%%%%%%%%%%%%%%%%%%%%%%%%%%%%%%%%%%%%%%%%%%%%%%
\subsubsection{Solutions with a sextic self-interaction term, $\beta > 0$}
%%%%%%%%%%%%%%%%%%%%%%%%%%%%%%%%%%%%%%%%%%%%%%%%%%%%%%%%%%%%%%%%%%%%%%%%%%%%%%%%
In what follows,
to simplify the picture,
we shall assume $\beta>1/4$,
such that
the potential $U(\psi^2)$ (as given by (\ref{U}))
is strictly positive (assuming $\psi^2>0$, which, as we shall see, is not necessarily the case).

Since no  exact solutions are known, the solitons are constructed numerically.
The profile of typical solutions are shown in Figures
\ref{scalar-flat}-\ref{Proca-flat}
(left panels).
One notices that
in the Proca case,
$\psi^2={\cal A}^2$
is negative for small enough $r$,
while
 $\psi^2= i \overline{\Psi}^{(A)}  \Psi^{(A)} $
is positive in the Dirac case.

%\newpage
%%%%%%%%%%%%%%%%%%%%%%%%%%%%%%%%%%%%%%%%%%%%%%%
 \begin{figure}[h!]
\begin{center}
\includegraphics[width=0.495\textwidth]{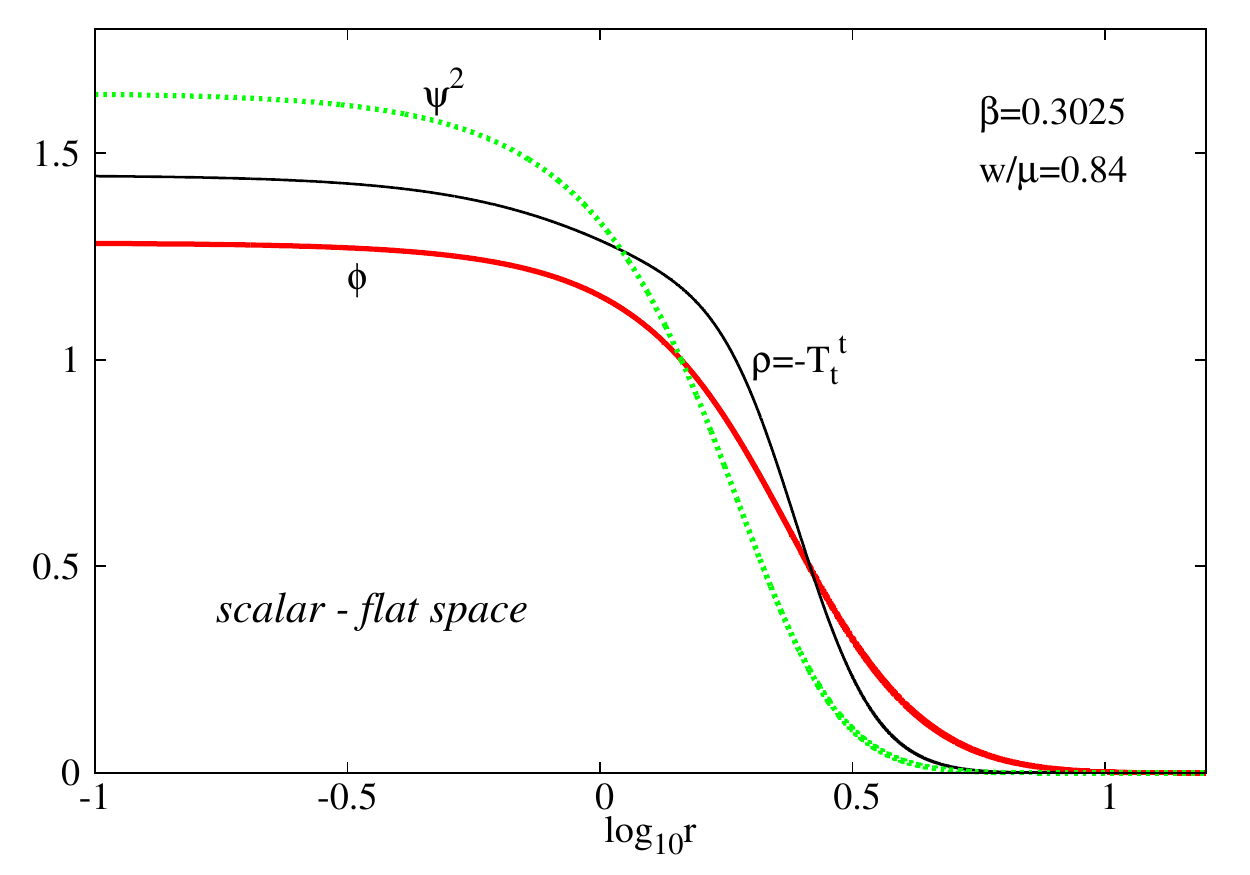}
\includegraphics[width=0.495\textwidth]{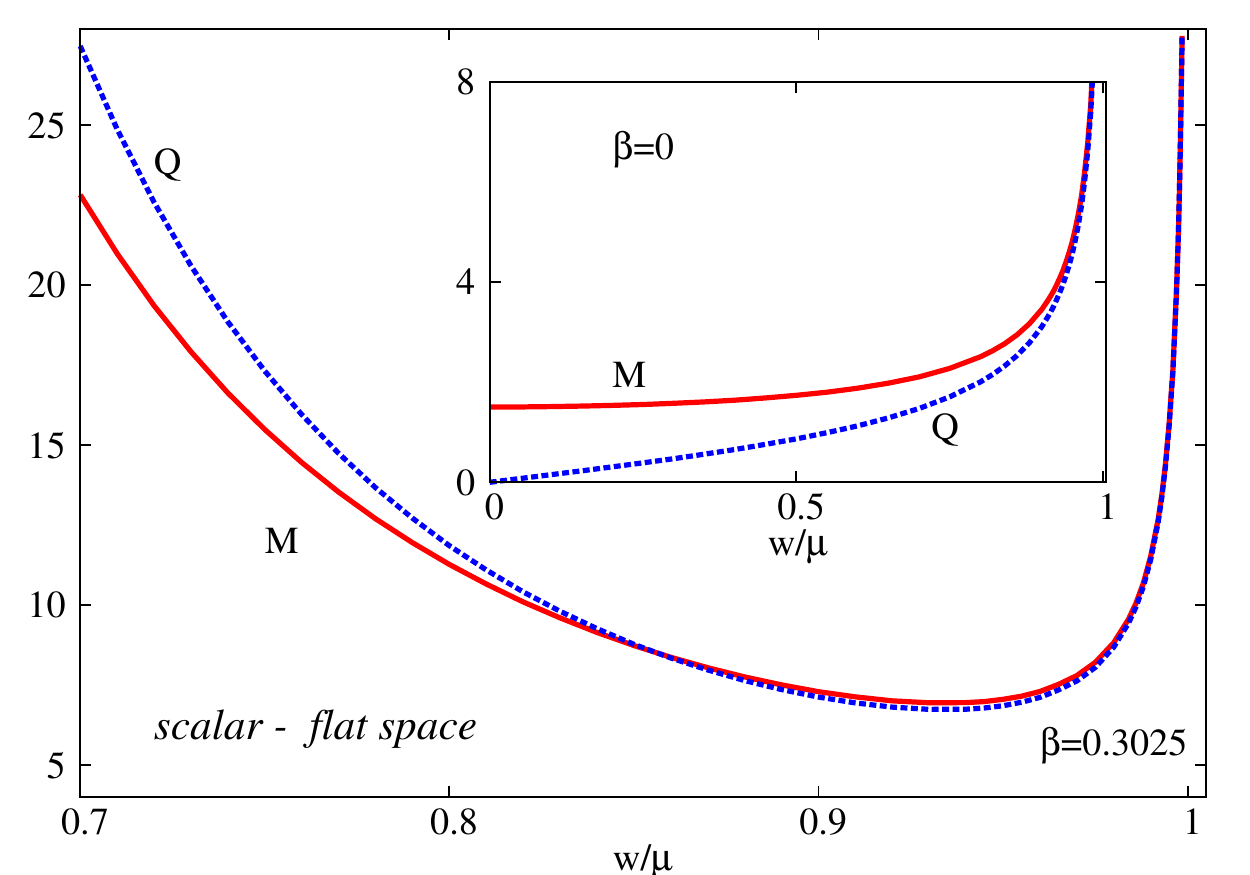}
%\\
%\includegraphics[width=0.495\textwidth]{flat-Dirac.eps}
%\includegraphics[width=0.495\textwidth]{Dirac.eps}
\caption{\small{
 {\it Left}: The radial profile of a typical non-gravitating scalar soliton.
{\it Right}:
The mass and Noether charge are shown $vs.$ the scalar field frequency
 for the fundamental family of non-gravitating scalar solitons.
}}
\label{scalar-flat}
\end{center}
\end{figure}
%%%%%%%%%%%%%%%%%%%%%%%%%%%%%%%%%%%%%%%%%%%%%%%

%%%%%%%%%%%%%%%%%%%%%%%%%%%%%%%%%%%%%%%%%%%%%%%
 \begin{figure}[h!]
\begin{center}
\includegraphics[width=0.495\textwidth]{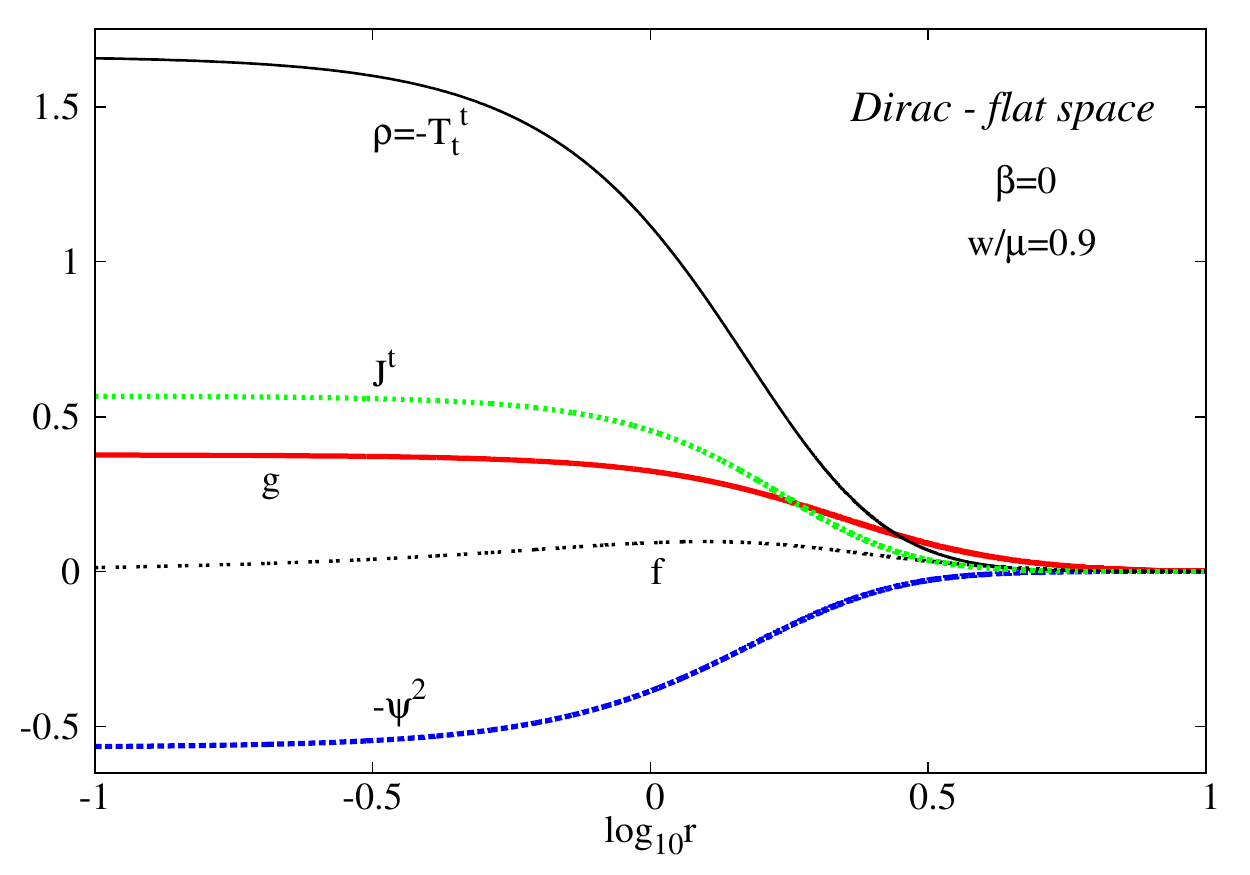}
\includegraphics[width=0.495\textwidth]{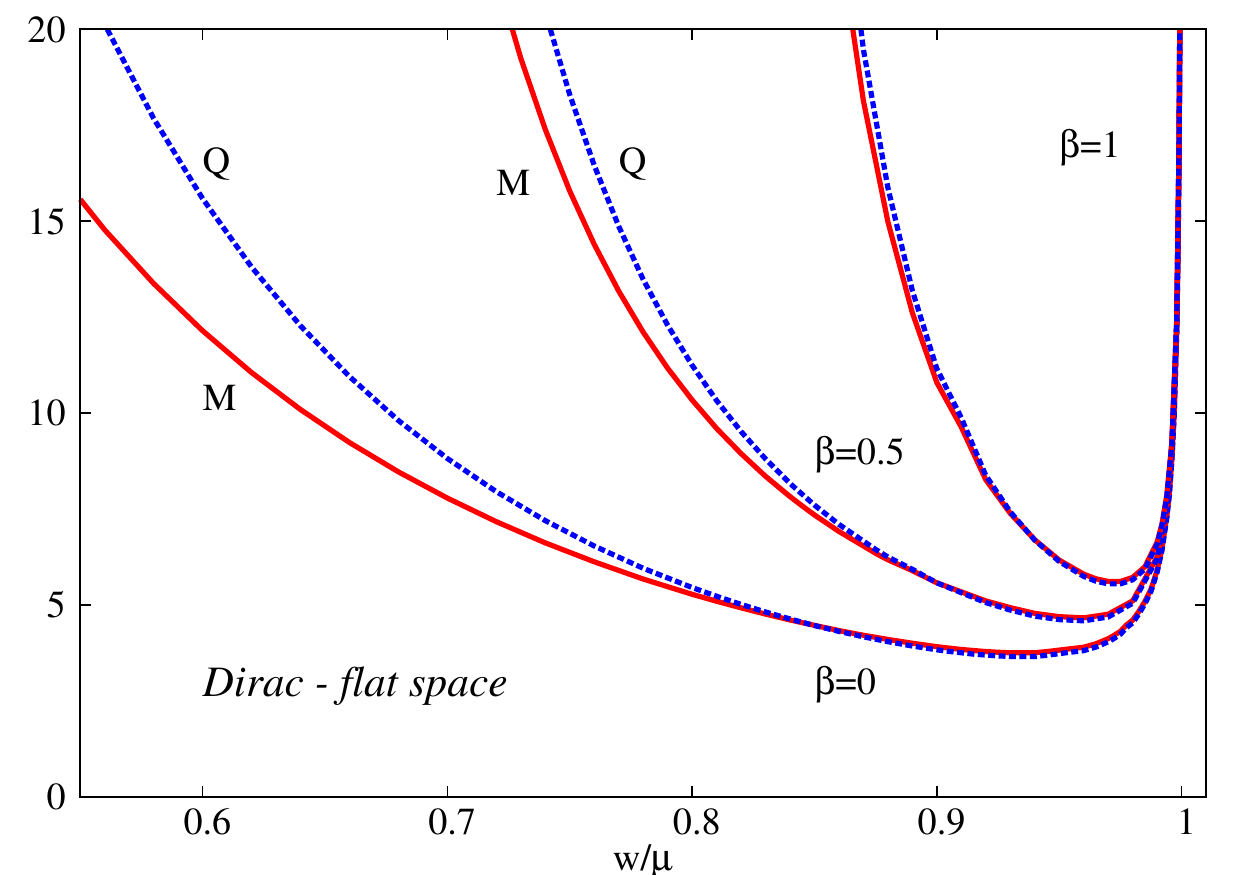}
%\\
%\includegraphics[width=0.495\textwidth]{flat-Dirac.eps}
%\includegraphics[width=0.495\textwidth]{Dirac.eps}
\caption{\small{
Same as Fig. \ref{scalar-flat} for Dirac stars.
Note that the single particle condition, $Q=1$, is not imposed here.
}}
\label{Dirac-flat}
\end{center}
\end{figure}
%%%%%%%%%%%%%%%%%%%%%%%%%%%%%%%%%%%%%%%%%%%%%%%

%%%%%%%%%%%%%%%%%%%%%%%%%%%%%%%%%%%%%%%%%%%%%%%
 \begin{figure}[h!]
\begin{center}
\includegraphics[width=0.495\textwidth]{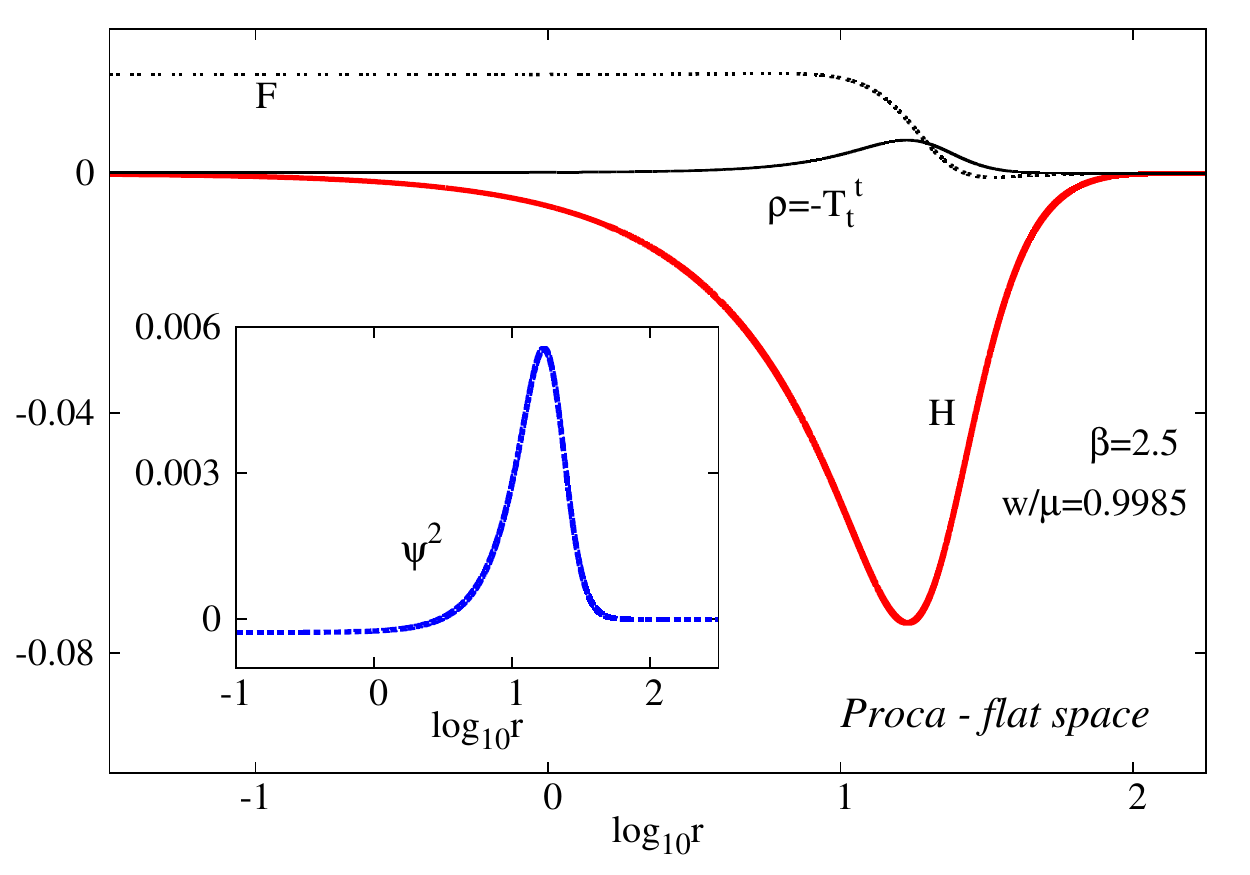}
\includegraphics[width=0.495\textwidth]{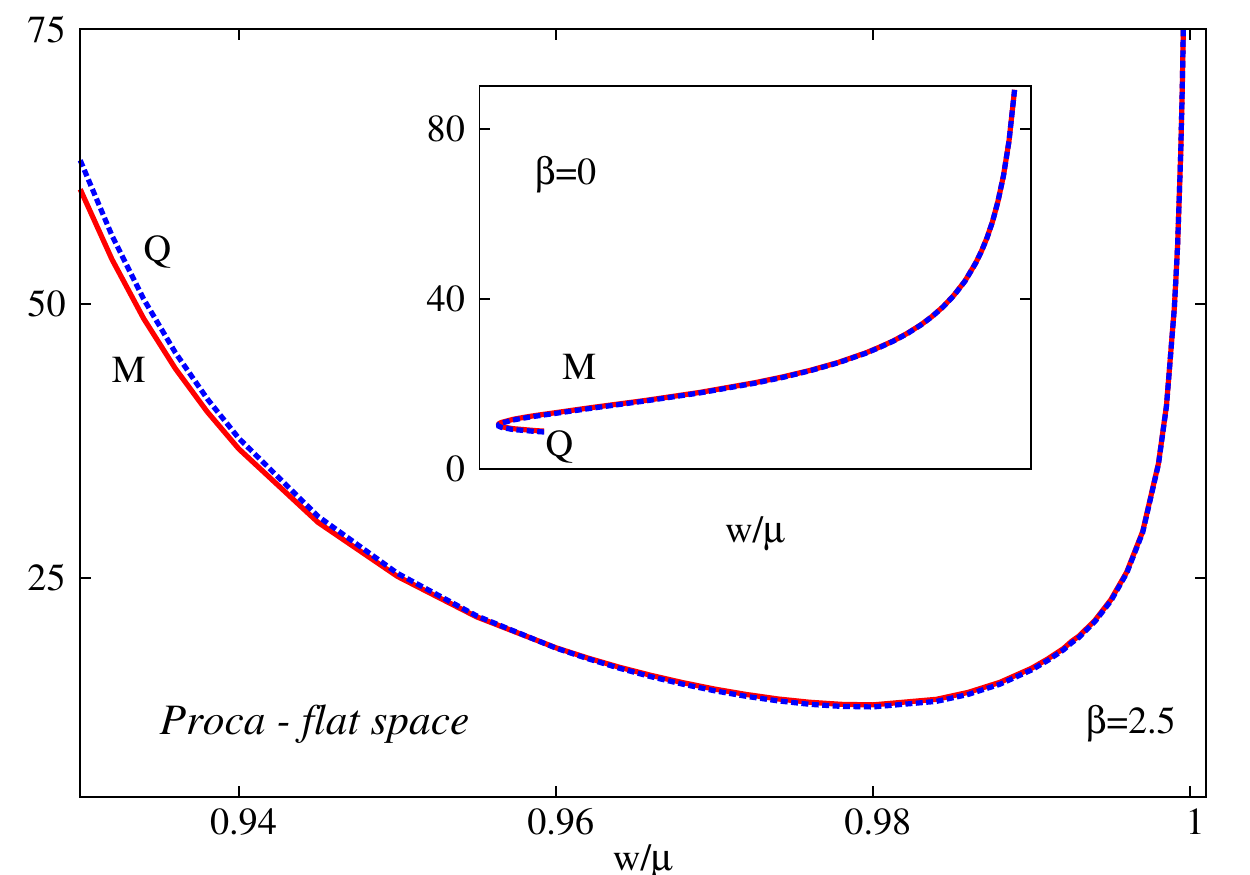}
%\\
%\includegraphics[width=0.495\textwidth]{flat-Dirac.eps}
%\includegraphics[width=0.495\textwidth]{Dirac.eps}
\caption{\small{
Same as Fig. \ref{scalar-flat} for Proca stars.
}}
\label{Proca-flat}
\end{center}
\end{figure}
%%%%%%%%%%%%%%%%%%%%%%%%%%%%%%%%%%%%%%%%%%%%%%%

The numerical results indicate that the $s=0,1/2,1$ flat spacetime
solitons follow the same pattern, which can be summarized
as follows.
First, in   all cases, the solutions exist only in a certain frequency range, $w_{\rm min} < w < w_{\rm max}=\mu$,
with $w_{\rm min}$ determined by $\beta$.
Second,
the solutions with $w <\mu $ decay exponentially in the far field, with no radiation, as
$\psi \sim e^{-\sqrt{\mu^2-w^2}r}$.
Finally, the most interesting qualitative feature is perhaps the existence {\it in all cases} of a mass gap.
That is, at a critical value of the frequency, both  the mass and the charge of the solutions assume
their minimal (nonzero)  value, from where they monotonically increase towards both limiting values of the frequency
(see Figures  \ref{scalar-flat}-\ref{Proca-flat} (right panels)).
 Considering
the mass of the solutions as a function of the charge, there are thus two branches, merging  at
the minimal charge/mass.
One expects that a subset of solutions with $M$ smaller than the mass of
$Q$ free particles (bosons or fermions) may be stable, which is possible along the lower frequency branch.

%%%%%%%%%%%%%%%%%%%%%%%%%%%%%%%%%%%%%%%%%%%%%%%%%%%%%%%%%%%%%%%%%%%%%%%%%%%%%%%%%%
\subsubsection{Solutions without a sextic self-interaction term, $\beta=0$}
%%%%%%%%%%%%%%%%%%%%%%%%%%%%%%%%%%%%%%%%%%%%%%%%%%%%%%%%%%%%%%%%%%%%%%%%%%%%%%%%%%

This case is likely physically less relevant (since the potential $U$ is not positive definite);
however, it possesses some interesting properties which depend, to some extent, on the spin of the field.

Starting with the scalar case,
some numerical results are displayed in Fig. \ref{scalar-flat}
(right panel, inset).
Note that all these solutions are unstable, since $M>Q$, and thus they have excess energy.
We remark that solutions with a negative energy density region, $\rho<0$,  found on a small $r-$interval,
are found for small enough $w$
(although $M>0$ always).

In fact,
the equation for the scalar field  has an interesting form,
the frequency parameter being  irrelevant.
After using the alternative rescaling
\begin{eqnarray}
r\to r/\sqrt{ \mu^2-w^2},~~~\phi \to \phi\sqrt{\frac{\mu^2 -w^2}{ \lambda}},
\end{eqnarray}
one can see that no free parameter exists in this case.
The scalar field satisfies the kink-like equation
\begin{eqnarray}
\label{coleman}
\frac{1}{r^2}\frac{d}{dr} \left(r^2 \frac{d \phi}{dr} \right)=\phi-2\phi^3,
\end{eqnarray}
which does not seem to possess an exact solution\footnote{
Eq. (\ref{coleman}) is discussed by many authors,
being relevant for the issue of false vacuum decay \cite{Coleman:1977py}
-- $e.g.$ an existence proof can be found in Ref. \cite{Coleman:1977th}.}.
Also, one can show that the following relation holds
\begin{eqnarray}
M(w;\mu,\lambda)= \frac{M(w=0;\mu,\lambda)}{\sqrt{ \mu^2-w^2}},
\end{eqnarray}
with $M(w=0;1,1) \simeq 1.503$.

\medskip
The pattern exhibited by the Proca solutions without a sextic term
is different, being displayed in  Fig. \ref{Proca-flat}.
As found in Ref. \cite{Brihaye:2017inn},
the limit $w=0$
is not approached in this case.
Instead, a minimal value of $w$ is reached,
with a backbending towards a critical solution possessing finite charges.
Since  $M>Q$, one expects
all these solutions to be unstable.
Also,  the mass is still positive, $M>0$,
and we did not notice the existence of negative energy
densities (although we could not find an analytical argument
to show that $\rho>0$).

\medskip

Finally, the pattern of Dirac solitons without a sextic self-interaction
(which was the original case in the pioneering work by  Soler \cite{Soler:1970xp} )
seems to be similar to the one found in the $\beta>0$ case,
see Figure \ref{Dirac-flat} (right panel).
In particular, both $M$ and $Q$ still diverge at the limits
of the $w$-interval.

%%%%%%%%%%%%%%%%%%%%%%%%%%%%%%%%%%%%%%%%%%%
\section{Including the gravity effects}
%%%%%%%%%%%%%%%%%%%%%%%%%%%%%%%%%%%%%%%%%%%

%%%%%%%%%%%%%%%%%%%%%%%%%%%%%%%%%%%%%%%%%%%
\subsection{The boundary conditions}
%%%%%%%%%%%%%%%%%%%%%%%%%%%%%%%%%%%%%%%%%%%
The boundary conditions satisfied by the  metric functions at the origin are
\begin{eqnarray}
m(0)=0,~~\sigma(0)=\sigma_0,
\end{eqnarray}
while at infinity  one imposes
\begin{eqnarray}
m(\infty)=M,~~\sigma(\infty)=1,
\end{eqnarray}
with  $\sigma(0)$, $M$  numbers fixed by numerics.
%Note that the above conditions for $\sigma$
%fixes the symmetry (\ref{s3}) of the system.
The matter functions should vanish as $r\to \infty$
\begin{eqnarray}
\phi(\infty)=f(\infty)=g(\infty)=F(\infty)=H(\infty)=0,
\end{eqnarray}
while the  boundary conditions at the origin
are
\begin{eqnarray}
\nonumber~
&&{\rm scalar:}~~
\frac{d\phi(r)}{dr}\bigg|_{r=0}=0,
\\
&&{\rm Dirac:}~~f(0)=0,~~\frac{dg(r)}{dr}\bigg|_{r=0}=0,
\\
\nonumber~
&&{\rm Proca:}~~\frac{dF(r)}{dr}\bigg|_{r=0}=0,~~H(0)=0.
\end{eqnarray}
Let us  mention that
in each case, one can construct an approximate form of the solutions both at $r=0$ and at infinity,
compatible with the boundary conditions above.

%%%%%%%%%%%%%%%%%%%%%%%%%%%%%%%%%%%%%%%%%%%
%\subsection{Numerical results}
%%%%%%%%%%%%%%%%%%%%%%%%%%%%%%%%%%%%%%%%%%%

%%%%%%%%%%%%%%%%%%%%%%%%%%%%%%%%%%%%%%%%%%%%%%%%%%
\subsection{Virial identities }
%%%%%%%%%%%%%%%%%%%%%%%%%%%%%%%%%%%%%%%%%%%%%%%%%%

The reduced action
(\ref{eff-action})
 allows us to
  prove that the solutions satisfy the  (simple enough) interesting virial identities,
	which generalize for a curved geometry Deser's relations (\ref{virial-scalar-flat}), (\ref{virial-Dirac-flat}, (\ref{virial-Proca-flat}).
	
Following~\cite{HS,Heusler:1996ft} (wherein generalizations for a curved geometry of Derrick's argument for flat spacetime \cite{Derrick:1964ww} were established),  assume the existence of a  solution described by $ m(r)$, $\sigma(r)$ and $\psi(r)$
(with $\psi =\{\phi;f,g; F,H \}$ the matter fields)
 with suitable boundary conditions at
$r=0$ and at infinity.
 Then each member of the 1-parameter family
$
m_\Lambda(r) \equiv m( \Lambda r),
$
$
\sigma_{\Lambda} (r) \equiv \sigma ( \Lambda r),
$
and
$
{\psi}_{\Lambda} (r) \equiv {\psi} ( \Lambda r),
$
assumes the same boundary values at $r=0$ and at $r=\infty$, and the action
$S_{\Lambda} \equiv S_{eff}[m_{\Lambda}, \sigma_{\Lambda},{  \psi}_{\Lambda}]$
must have a critical point at $\Lambda=1$,
$i.e.$
$[dS/d\Lambda]_{\Lambda=1}=0$.
Thus,  the  putative  solution must satisfy the following  virial relations
(here it is useful to work with unscaled variables, $i.e.$ before
considering the transformations (s2), (s3)):
\begin{itemize}
\item
scalar field:
\end{itemize}
\begin{eqnarray}
\label{virial-scalar}
 \int_0^\infty dr~r^2 \sigma
\left[
 \phi'^2+\frac{  w^2(1-4 N)\phi^2}{N^2\sigma^2}+3  U
\right]=0 ~;
\end{eqnarray}

\begin{itemize}
\item
Dirac field:
\end{itemize}
\begin{eqnarray}
\label{virial-Dirac}
 \int_0^\infty dr~r^2\sigma
\left [
 \sqrt{H}(g f'-fg')\left(2+\frac{m}{r N}\right)
+\frac{4 fg}{r}
-\frac{w(f^2+g^2)}{\sigma \sqrt{N}}\left(3-\frac{m}{rN}\right)-3U
\right ]=0~;
\end{eqnarray}
%where
%\begin{eqnarray}
%\label{virial-Dirac2}
% U=\mu(f^2-g^2)+2\lambda (f^2-g^2)^2+16 \nu (f^2-g^2)^3;
%\end{eqnarray}

 \begin{itemize}
\item
Proca field:
\end{itemize}
\begin{eqnarray}
\label{virial-Proca}
 \int_0^\infty dr~r^2 \sigma
\bigg [
 \mu^2
\left(
{\cal A}^2\left(4-\frac{1}{N} \right)+2 H^2(1-N)
\right)
+2\lambda {\cal A}^2
\left(
{\cal A}^2 \left(5-\frac{2}{N}\right)
+4H^2(1-N)
\right)
\\
\nonumber
+ 6 \nu {\cal A}^4
\left(
 {\cal A}^2 \left(2-\frac{1}{N}\right)+2 H^2(1-N)
+4G^2(1-N)
\right)
-\frac{(wH-F')(3wH-F')}{\sigma^2}
\bigg ]=0 .
\end{eqnarray}
This relation can be further simplified after using the Proca field equations to yield
\begin{eqnarray}
\label{virial-Proca2s}
  \int_0^\infty dr~r^2 \sigma
\left [
\mu^2\left(4-\frac{1}{N}\right){\cal A}^2+
2\lambda \left(5-\frac{2}{N}\right){\cal A}^4
+6\nu \left(2-\frac{1}{N}\right){\cal A}^6
\right]=
\\
\nonumber
 2 \int_0^\infty dr~r^2 \sigma H^2
( \mu^2+4 \lambda {\cal A}^2+6\nu  {\cal A}^4)
\left [
 2N-1
+\frac{N^2\sigma^2}{2w^2}
( \mu^2+4 \lambda {\cal A}^2+6\nu {\cal A}^4)
\right]
\end{eqnarray}

These expressions are not transparent, with
the effects of gravity and self-interaction being mixed.
As such their main use  is to test the accuracy of the numerical results.
However, the situation changes
once we set $\lambda=\nu=0$ ($i.e.$ no self-interaction).
Then one can see that the solutions
are supported by an harmonic time dependence of
the matter fields,
$w\neq 0$, except for the Dirac case, where we could not prove a similar result.

%%%%%%%%%%%%%%%%%%%%%%%%%%%%%%%%%%%%%%%%%%%
\subsection{General features}
%%%%%%%%%%%%%%%%%%%%%%%%%%%%%%%%%%%%%%%%%%%

The flat space solitons can be generalized
to curved spacetime.
The presence of higher order self-interacting terms in the potential
is not crucial for the existence of gravitating solutions.
However, they affect some of their quantitative features.

\medskip
A common pattern emerges again,
the basic generic properties of the gravitating solutions being summarized as follows.
First, the solutions are topologically trivial, with $0\leqslant r<\infty$.
They possess no horizon; the size of
the $S^2$-sector of the metric shrinks to zero as $r\to 0$.
At infinity, a Minkowski background is approached.
Second, perhaps the most interesting feature
is that gravity regularizes the divergencies of the mass and charge
found in the probe case at the limit(s) of the $w$-interval.
In particular, this regularization  implies no mass gap exists for gravitating configurations: $M,Q \to 0$ as $w\to \mu$.
Also, assuming the existence of a sextic self-intertion term, with $\beta>1/4$,
in all cases, the family of solutions
describes a continuous curve in a mass $M$ (or charge $Q$), $vs.$  frequency, $w$, diagram.
This curve starts from $M=Q=0$ for $w=\mu$, in which limit the fields becomes very diluted and the solution trivializes.
 At some intermediate frequency, a maximal mass is attained (which may be a global, or only local, maximum, depending on $\alpha,\beta)$.

The effects of the quartic and sextic self-interacting terms in
the potential
(\ref{U})
become irrelevant for
large enough $\alpha$.
For example, the $\alpha=10$ curves displayed in
Figures \ref{scalar-grav}-\ref{Proca-grav}
are well approximated by the
corresponding Einstein-scalar/Dirac/Proca results
with $\beta=\nu=0$,
the maximal relative difference being of only a few percent (towards the critical value of frequency).

%%%%%%%%%%%%%%%%%%%%%%%%%%%%%%%%%%%%%%%%%%%%%%%
 \begin{figure}[h!]
\begin{center}
\includegraphics[width=0.495\textwidth]{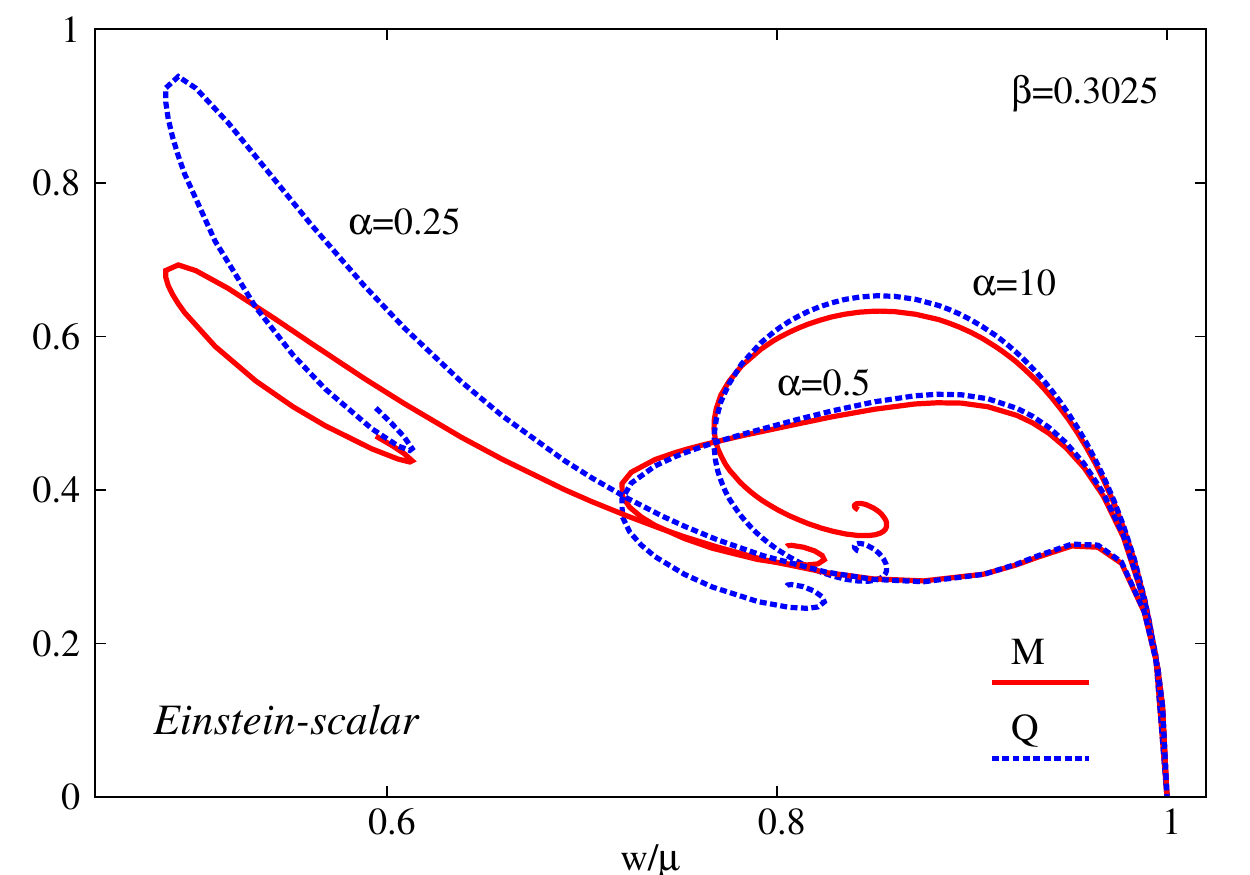}
\includegraphics[width=0.495\textwidth]{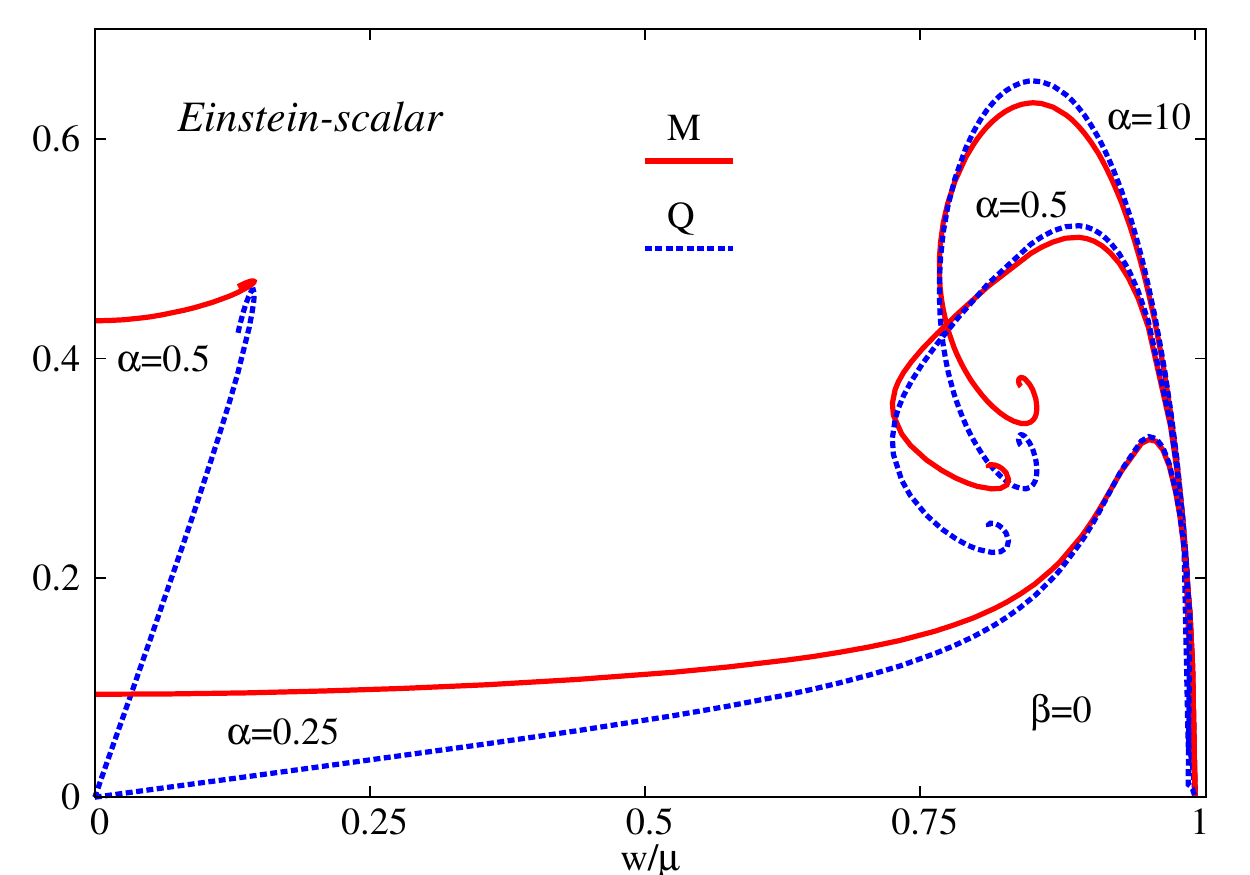}
\caption{\small{
 % (Left panel)
ADM mass and Noether charge of the gravitating scalar boson stars  $vs.$ the scalar field frequency  for
families of
solutions with three different values of the coupling constant $\alpha$.
The solutions in the right panel do not possess a  sextic self-interacting term.
}}
\label{scalar-grav}
\end{center}
\end{figure}
%%%%%%%%%%%%%%%%%%%%%%%%%%%%%%%%%%%%%%%%%%%%%%%

%%%%%%%%%%%%%%%%%%%%%%%%%%%%%%%%%%%%%%%%%%%%%%%
 \begin{figure}[h!]
\begin{center}
\includegraphics[width=0.495\textwidth]{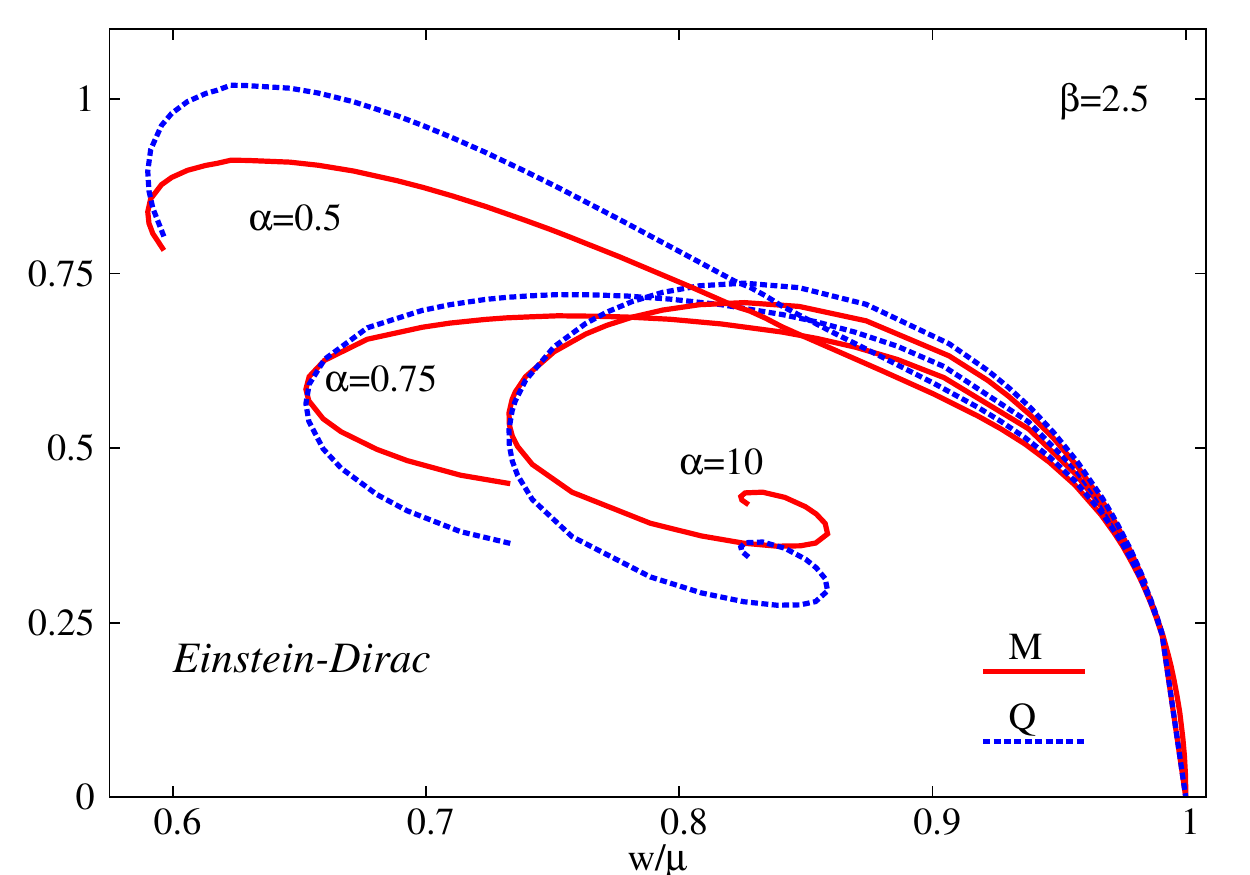}
\includegraphics[width=0.495\textwidth]{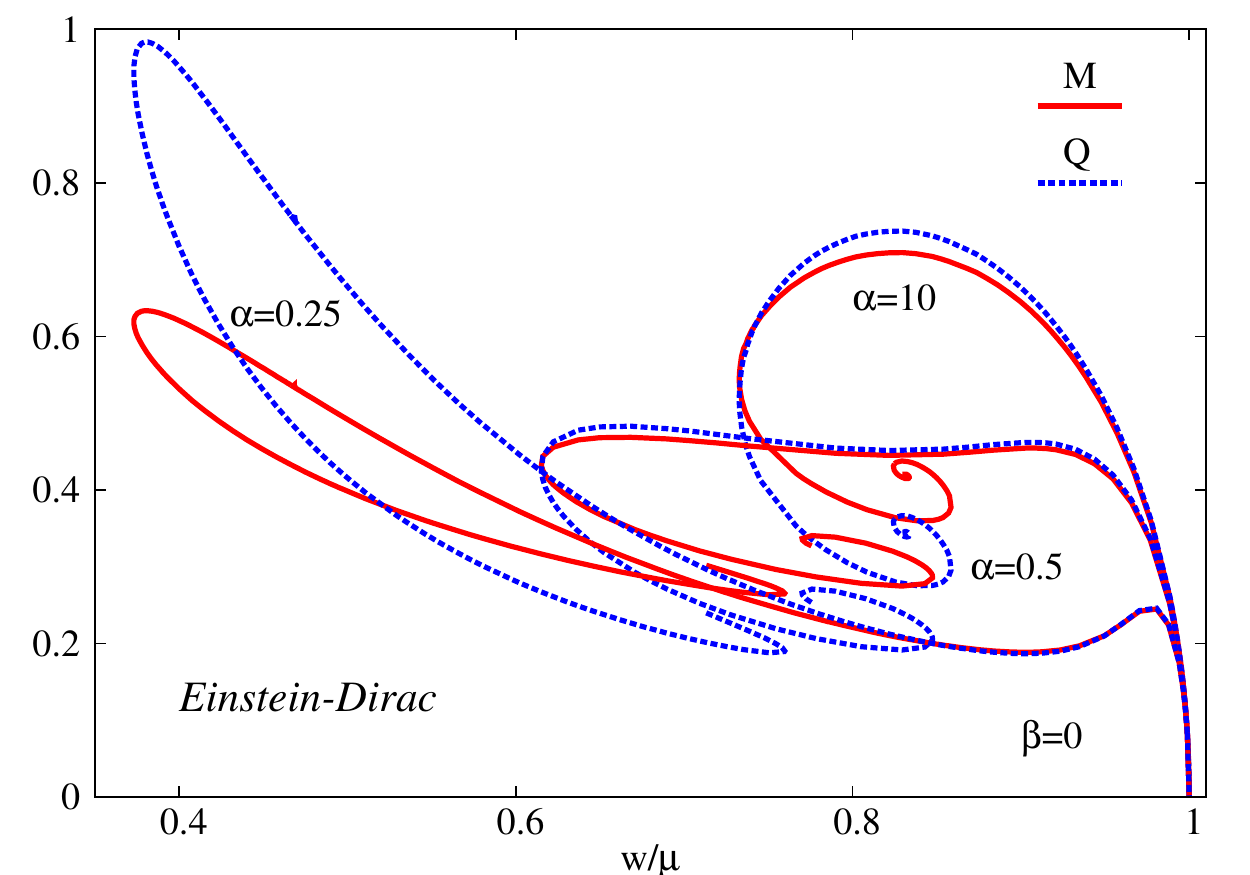}
%\\
%\includegraphics[width=0.495\textwidth]{flat-Dirac.eps}
%\includegraphics[width=0.495\textwidth]{Dirac.eps}
\caption{\small{
Same as Fig. \ref{scalar-grav} for Dirac stars.
The single particle condition, $Q=1$, is not imposed here.
}}
\label{Dirac-grav}
\end{center}
\end{figure}
%%%%%%%%%%%%%%%%%%%%%%%%%%%%%%%%%%%%%%%%%%%%%%%

\medskip

One can also identify also some specific features, as follows.
The   scalar solutions  with $\beta>1/4$ ($i.e.$ a positive potential, $U>0$)
and the Proca starts with large $\alpha$
describe a spiral in a $(M,w)$-diagram.
As in similar cases,  likely, these spirals approach, at their centre, a critical singular solution.
The Dirac solutions with $\beta=0$ also describe a spiral.
For general
  Proca solutions with $\beta=0$
and   Dirac solutions with $\beta \neq 0$,
the
$(M,w)$-curves appear to end
in a critical configuration
before describing a spiral.
In the Proca case,
this feature is discussed in \cite{Brihaye:2017inn}.

The scalar field solutions with $\beta=0$
possess a more complicated pattern.
For small $\alpha$, they possess a static limit.
Moreover, two disconnected branches of solutions exist for some
intermediate range
of $\alpha$ - $e.g.$ $\alpha=0.5$  in Figure \ref{scalar-grav} (right panel).
In addition to the familiar spiral starting at $w/\mu=1$
and ending for some critical nonzero $w=w_c$,
one finds a secondary set,
 extending from $w=0$ to some maximal
value of $w<w_c$.

Finally, let us mention the existence of another interesting possibility, with quartic interaction
only and $\lambda<0$, in which case
no flat spacetime solitons  are found.
The corresponding solutions were discussed in
\cite{Colpi:1986ye}
for a scalar field,
and in
\cite{Minamitsuji:2018kof}
for a Proca field.
No similar study exists so far for a fermionic field.
In the bosonic case, perhaps the most interesting
feature is that their maximal mass is proportional with
$\sqrt{|\lambda|}$, and thus can increase dramatically.

%%%%%%%%%%%%%%%%%%%%%%%%%%%%%%%%%%%%%%%%%%%%%%%
 \begin{figure}[h!]
\begin{center}
\includegraphics[width=0.495\textwidth]{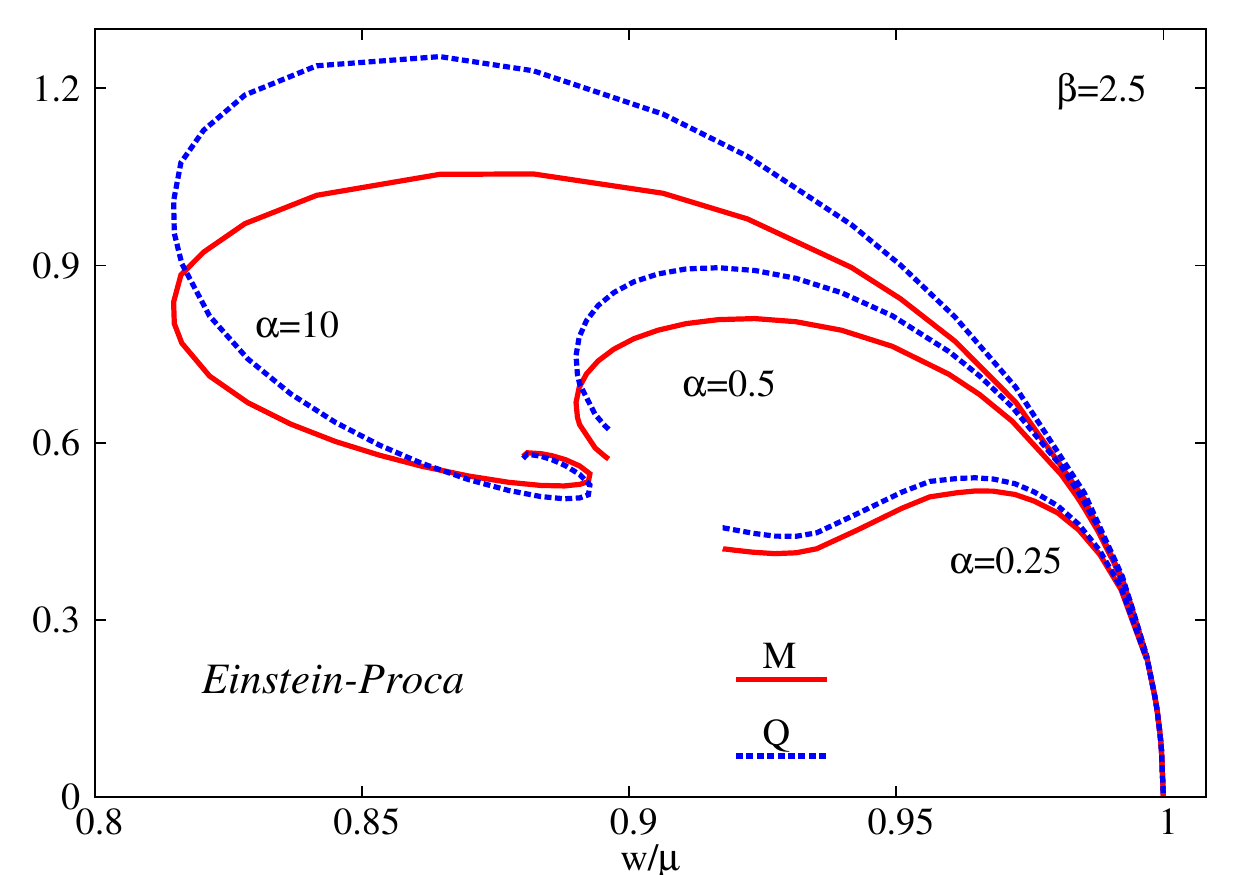}
\includegraphics[width=0.495\textwidth]{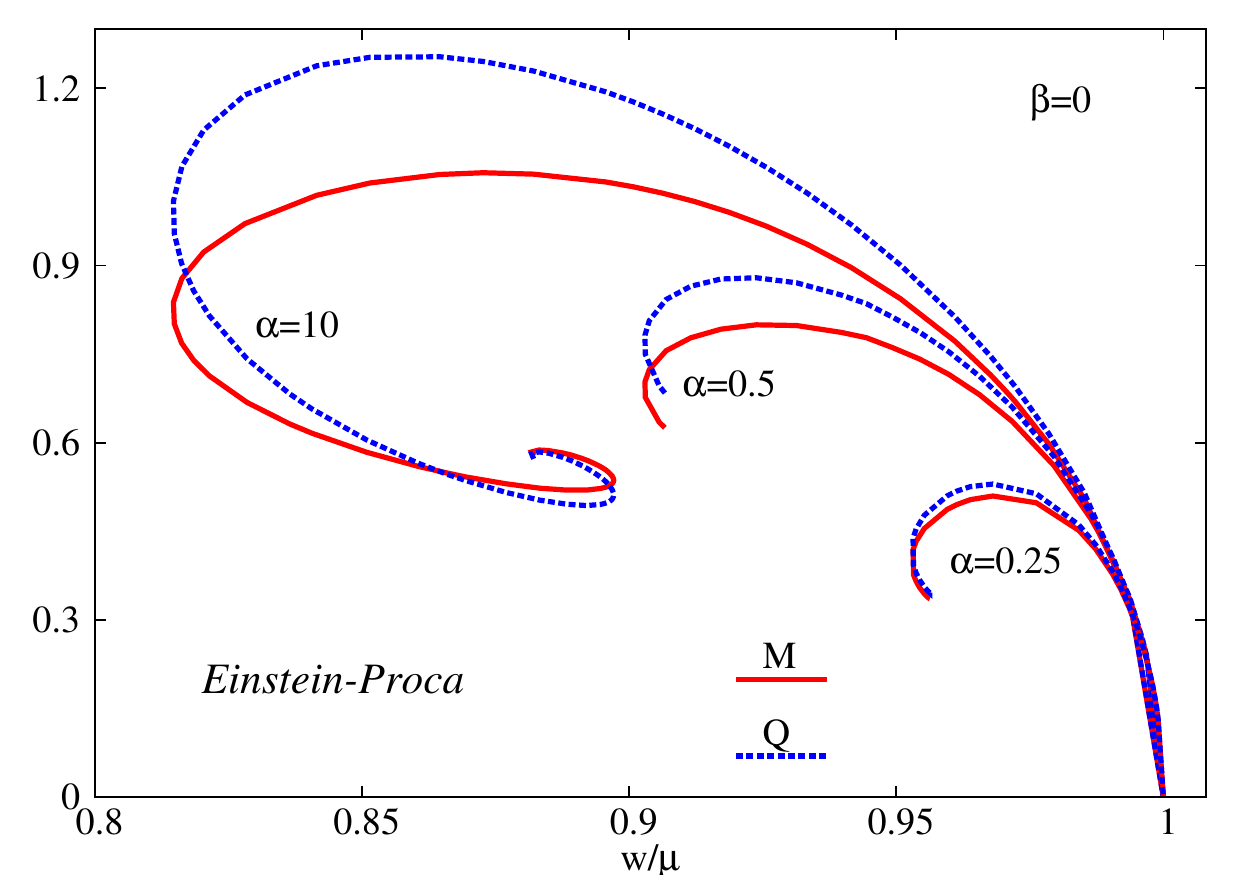}
\caption{\small{
Same as Fig. \ref{scalar-grav} for Proca stars.
}}
\label{Proca-grav}
\end{center}
\end{figure}

%%%%%%%%%%%%%%%%%%%%%%%%%%%%%%%%%%%%%%%%%%%%%%%

%%%%%%%%%%%%%%%%%%%%%%%%%%%%%%%%%%%%%%%%%%%%%%%%%%
\section{Other aspects}
%%%%%%%%%%%%%%%%%%%%%%%%%%%%%%%%%%%%%%%%%%%%%%%%%%

%%%%%%%%%%%%%%%%%%%%%%%%%%%%%%%%%%%%%%%%%%%%%%%%%%
\subsection{No hair results }
%%%%%%%%%%%%%%%%%%%%%%%%%%%%%%%%%%%%%%%%%%%%%%%%%%

It is interesting to inquire if the solutions above
may allow the existence of a black hole horizon, inside.
Indeed, this is a well known feature found for a variety of
other solitons, see $e.g.$ the review work \cite{Volkov:1998cc}.
However, this is not the case for the (relatively) simpler solutions in
this work.
%, a feature which appears to be
%connected with the existence of a Noether charge.

The situation can be summarized as follows:
\begin{itemize}
\item
{\bf Scalar case}
\\
 Pe\~{n}a and Sudarsky established a no-scalar-hair theorem,
ruling out a  class of spherically symmetric
 BHs with scalar hair \cite{Pena:1997cy}.
Their proof covers also the case of the potential (\ref{U})
considered in this work, and
still holds
  if the hair has the harmonic time-dependence we consider.

\item
{\bf Dirac case}
\\
No hair results for the Einstein-Dirac case with a massive spinor
(no self-interaction ) were proposed in
\cite{Finster:1998ak,Finster:1998ju}.
The nonexistence of stationary
states for the nonlinear Dirac equation with a quartic self-interaction
 on the Schwarzschild metric
has been proven in
\cite{BachelotMotet:1998ij}.

\item
{\bf Proca case}
\\
A no hair theorem has been  proven in
\cite{Herdeiro:2016tmi}
for a massive, non-selfinterating Proca field.
In appendix B, we generalize it
for an arbitrary Proca potential $U({\cal A}^2)$.

\end{itemize}

%%%%%%%%%%%%%%%%%%%%%%%%%%%%%%%%%%%%%%%%%%%%%%%%%%%%%%%%%%%%%%%%
 \subsection{The issue of particle numbers: bosons vs. fermions}
%%%%%%%%%%%%%%%%%%%%%%%%%%%%%%%%%%%%%%%%%%%%%%%%%%%%%%%%%%%%%%%%
In all results displayed above,
we have treated equally the bosonic and fermionic fields.
However, while the classical treatment of
the Dirac equation is mathematically  justified\footnote{A discussion on classical spinors and  their possible physical justification can be found, $e.g.$,  in Ref.  \cite{ArmendarizPicon:2003qk}.}, physically,
its fermionic nature should be imposed at the level of the occupation number: at most a single,
particle, in accordance to Pauli's exclusion principle.

Similarly to the non-self-interacting case
discussed in \cite{Herdeiro:2017fhv},
the one particle condition
is imposed as follows.
%by using the transformation (\ref{nc1}), (\ref{nc2}).
%
Let us suppose we have a numerical solution with some values for the mass and charge
($M_{\rm (num)}$, $Q_{\rm (num)}$).
Then we can use the  symmetry (\ref{s1})-(\ref{s1-MQ})
in order to map it to a 'new' solution with
\begin{eqnarray}
 \bar Q_{\rm (num)}=Q_0,
\end{eqnarray}
where $Q_0>0$ is arbitrary.
This assumption fixes the value of the scaling parameter $c$,
\begin{eqnarray}
\label{nc1}
c=\sqrt{\frac{Q_{\rm (num)}}{Q_0}},
\end{eqnarray}
such that the $numerical$ mass of the 'new' solution will be
\cite{Herdeiro:2017fhv}
\begin{eqnarray}
\label{nc2}
 \bar M_{\rm (num)}= M_{\rm (num)}  \sqrt{\frac{Q_0}{ Q_{\rm (num)}}}~.
\end{eqnarray}
Note that the values of $w$, $\mu$, $\lambda$ and $\nu$
should also be scaled according to (\ref{s1}), (\ref{s1-dirac}).

\medskip
This transformation is used
to normalize the total charge of a fermion to one, $Q_0=1$.
With this condition\footnote{Note that in the Dirac  case we impose $Q=1$ for each spinor.
}
$Q=1$
the $(M,w)$-curves in
Fig.~\ref{Dirac-flat}  (right panels)
and
Fig. \ref{Dirac-grav}
are not sequences of solutions with fixed parameters in the potential and varying $Q$;
instead, they get mapped into
 sequences with fixed $Q$ and varying $\mu,\nu$.
Consequently, one is discussing \textit{a sequence of solution of different models}
(since $\mu$, $\nu$ are input parameters in the action).

%%%%%%%%%%%%%%%%%%%%%%%%%%%%%%%%%%%%%%%%%%%%%%%
 \begin{figure}[h!]
\begin{center}
\includegraphics[width=0.495\textwidth]{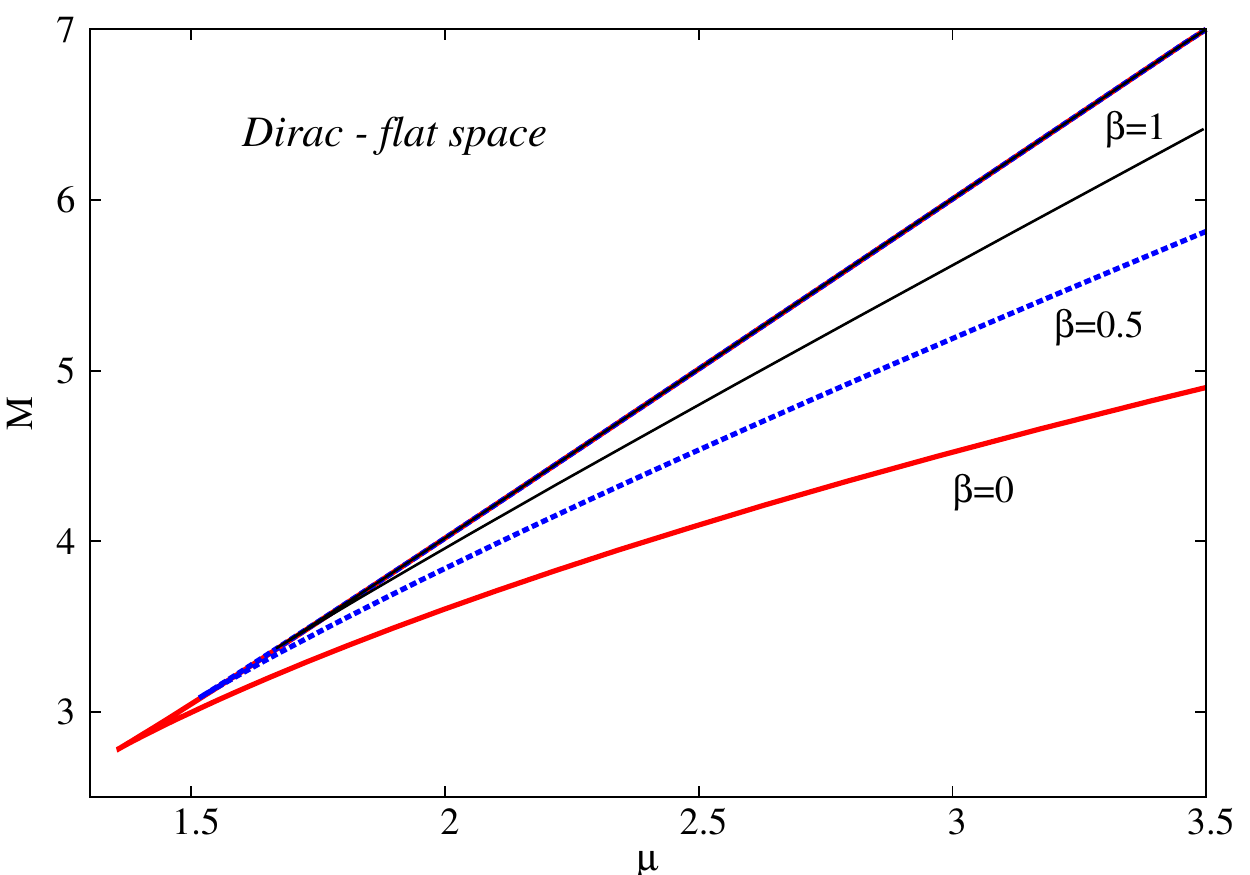}
\includegraphics[width=0.495\textwidth]{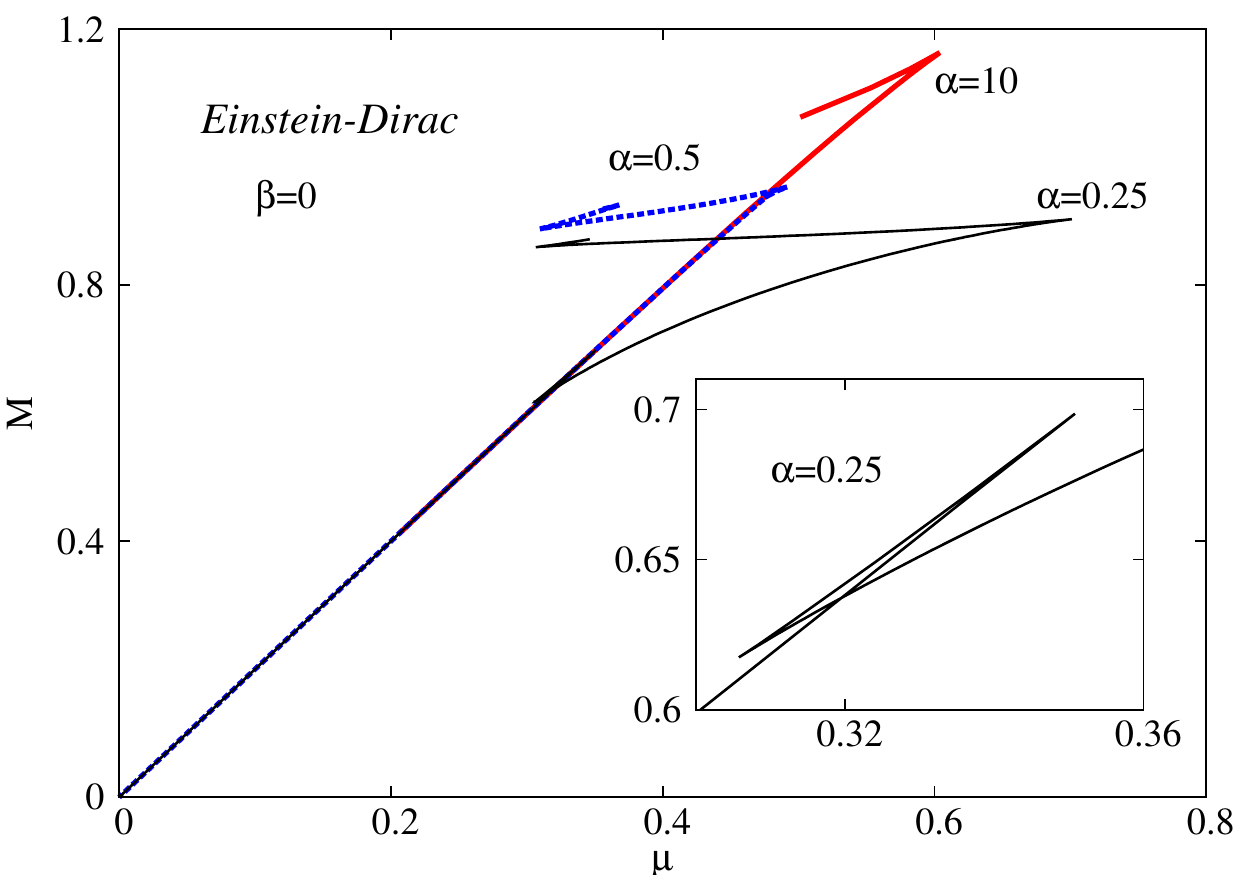}
%\\
%\includegraphics[width=0.495\textwidth]{flat-Dirac.eps}
%\includegraphics[width=0.495\textwidth]{Dirac.eps}
\caption{\small{
(Left panel)
Soliton mass $vs.$  the  mass of the elementary quanta of the field, for non-gravitating
solutions of the Dirac equations with three different values of $\beta$.
(Right panel)
 Same for the  gravitating solutions with several values of the coupling constant $\alpha$
and $\beta=0$.
The single particle condition, $Q = 1$, is imposed here.
}}
\label{figD1}
\end{center}
\end{figure}
%%%%%%%%%%%%%%%%%%%%%%%%%%%%%%%%%%%%%%%%%%%%%%%

The corresponding results are shown in Fig. \ref{figD1},
for both the probe limit case and including gravity effects.
An interesting feature here is that the mass of $Q=1$
non-gravitating
configuration can still take arbitrary large values.
However, as expected, gravity effects lead to a picture qualitatively similar to that found in
the $\lambda=\nu=0$
case \cite{Herdeiro:2017fhv}.
Again, both the total mass, $M$ and the mass of the field $\mu$ are
bounded  and never exceeds, roughly, $M_{Pl}$.

%%%%%%%%%%%%%%%%%%%%%%%%%%%%%%%%%%%%%%%%%%%
\section{Further  remarks. Conclusions}
%%%%%%%%%%%%%%%%%%%%%%%%%%%%%%%%%%%%%%%%%%%

The main purpose of this work was to provide a comparative analysis of three different types of solitonic
solutions of GR coupled with matter fields
of spin $s=0,1/2,1$.
A unified framework has been proposed,  analysing these cases side by side under a consistent set of notations and conventions.
Differently from the previous work \cite{Herdeiro:2017fhv},
 the matter fields herein
 are self-interacting, such that all three models
possess ($Q$-ball--like) solutions on the flat spacetime limit.
As such, a more complicated landscape of
gravitating
solutions is found,
with some new qualitative features as compared to the picture revealed in \cite{Herdeiro:2017fhv}.

Despite this fact, however,
our study shows that there is again a certain universality in the properties of the solitons, being to some extent independent of their (fundamental) spin.
As with linear matter fields, the basic ingredients are again:
${\bf i)}$ a harmonic time dependence of the matter fields;
 ${\bf ii)}$ complex field(s)/multiplets such that the energy-momentum is still real
and
  ${\bf iii)}$
the existence of of mass term as a trapping mechanism, creating  bound states with $w<\mu$.
Also, if one requires the presence of a well defined flat space limit,
then
${\bf iv)}$ the fields should possess at least a quartic self-interaction term.

\medskip
As an avenue for future work, it would be interesting to go beyond the case of spherical
symmetry and  consider a comparative study of axially symmetric, spinning solutions.
In the non-self-interacting case, this was the subject of the recent work \cite{Herdeiro:2019mbz},
where a common pattern was again revealed to exist.
The situation in the presence of self-interactions
is less studied;  only the scalar field case  has,  so far, been discussed  in the literature
\cite{Volkov:2002aj,Kleihaus:2005me}. Indeed, even flat space spinning solitons with spin $s=1/2,1$
are yet unreported in the literature.
Moreover,   even in the static case,
 new families of non-spherically symmetric solitons should exist,
generalizing for a  self-interacting potential (and possibly for a higher spin)
the $s=0$ multipolar boson stars recently reported in~\cite{Herdeiro:2020kvf}.

Finally, let us remark that
 that for a  bosonic field, $s=0,1$,
the no-hair theorems
(as reviewed in the previous Section)
 can be circumvented for an horizon that
rotates
synchronously with the field, leading to
BHs with scalar or Proca hair
\cite{Herdeiro:2014goa},
\cite{Herdeiro:2016tmi}.
 However,  this does not seem to be the case for fermions, regardless of the presence
of a  self-interacting potential.

%%%%%%%%%%%%%%%%%%%%%%%%%%%%%%%%%%%%%%%%%%%%%%%
\section*{Acknowledgements}
%%%%%%%%%%%%%%%%%%%%%%%%%%%%%%%%%%%%%%%%%%%%%%%

This  work  is  supported  by  the Center  for  Research  and  Development  in  Mathematics  and  Applications  (CIDMA)  through  the Portuguese Foundation for Science and Technology (FCT - Fundacao para a Ci\^encia e a Tecnologia), references UIDB/04106/2020 and UIDP/04106/2020 and by national funds (OE), through FCT, I.P., in the scope of the framework contract foreseen in the numbers 4, 5 and 6 of the article 23, of the Decree-Law 57/2016, of August 29, changed by Law 57/2017, of July 19.  We acknowledge support  from  the  projects  PTDC/FIS-OUT/28407/2017,  CERN/FIS-PAR/0027/2019 and PTDC/FIS-AST/3041/2020.   This work has further been supported by the European Union’s Horizon 2020 research and innovation (RISE) programme H2020-MSCA-RISE-2017 Grant No. FunFiCO-777740.
The authors would like to acknowledge networking support by the COST Action CA16104.
Computations were performed
at the Blafis cluster, in Aveiro University.

\appendix
\setcounter{equation}{0}

%%%%%%%%%%%%%%%%%%%%%%%%%%%%%%%%%%%%%%%%%%%%%%%%%%%%%%%%%%%%%%%%%%%%%%%%%%%%%%
\section{The Dirac field: conventions}
%%%%%%%%%%%%%%%%%%%%%%%%%%%%%%%%%%%%%%%%%%%%%%%%%%%%%%%%%%%%%%%%%%%%%
Since this case is more complicated, we shall include here the basic relations.

Following the framework
in \cite{Dolan:2015eua}, 
we consider  a general four dimensional metric $g_{\alpha \beta}$,
and  introduce a tetrad of vectors
\begin{eqnarray}
e_a^\alpha = \{e_0^\alpha, e_1^\alpha, e_2^\alpha, e_3^\alpha \},
\end{eqnarray}
 which we take to be an orthonormal basis,
$i.e.$,
\begin{eqnarray}
g_{\alpha \beta} e_a^\alpha e_b^\beta = \eta_{\alpha \beta},
\end{eqnarray}
 where
\begin{eqnarray}
\eta_{ab} = {\rm diag}(-1, 1, 1, 1).
\end{eqnarray}
Here,  Roman and Greek letters are used for tetrad and coordinate indices, respectively.
Roman indices are raised and lowered with $\eta_{ab}$. It follows that
\begin{eqnarray}
e^a_\alpha = \eta^{a b} g_{\alpha \beta} e_b^\beta~~~ {\rm and}~~~g_{\alp\bet} = \eta_{ab} e_{\alp}^a e_{\bet}^b.
\end{eqnarray}

The next step is to
  define two sets of $4 \times 4$ matrices $\gamma^{\alpha}$ and $\hat{\gamma}^{a}$ satisfying the anticommutation relations
\begin{eqnarray}
\label{gamanti}
\{ \gamma^\alpha, \gamma^\beta \}  = 2 g^{\alpha \beta} I_4 ,~~~
 \{ \hat{\gamma}^a, \hat{\gamma}^b \}  = 2 \eta^{ab} I_4,
\end{eqnarray}
(where, as usual, $\{A,B\} = A B + B A$).
 Note that the former set $\gamma^\alpha$ are functions of spacetime position,
whereas the latter $\hat{\gamma}^a$ have constant components. The two sets may be related with any orthonormal tetrad,
\beq
\gamma^\alpha =  e_a^\alpha \hat{\gamma}^a .
\eeq
 Matrix indices are raised/lowered in the standard way:
 $\hat{\gamma}_a = \eta_{ab} \hat{\gamma}^b$ and $\gamma_\alpha = g_{\alp \bet} \gamma^\bet$.

One uses the Weyl/chiral representation, in which
\begin{eqnarray}
\tilde{\gam}^0 = \begin{pmatrix} O & I \\ I & O \end{pmatrix}, \quad \tilde{\gam}^i = \begin{pmatrix} O & \sig_i \\ -\sig_i & O \end{pmatrix} , \quad \quad i = 1,2,3,
\end{eqnarray}
where $I$ is the $2\times2$ identity, $O$ is the $2 \times 2$ zero matrix, and $\sig_i$ are the Pauli matrices
\begin{eqnarray}
\sig_1 = \begin{pmatrix} 0 & 1 \\ 1 & 0 \end{pmatrix},
\quad
\sig_2 = \begin{pmatrix} 0 & -i \\ i & 0 \end{pmatrix},
\quad
\sig_3 = \begin{pmatrix} 1 & 0 \\ 0 & -1 \end{pmatrix} .
\end{eqnarray}
Then the matrices $\hat{\gamma}^{a}$
are defined as
\begin{eqnarray}
\hat{\gamma}^1 = i \tilde{\gamma}^3,~~
\hat{\gamma}^2 = i \tilde{\gamma}^1,~~
\hat{\gamma}^3 = i \tilde{\gamma}^2,~~
\hat{\gamma}^0 = i \tilde{\gam}^0.
\end{eqnarray}

The Dirac four-spinor is written as
\begin{eqnarray}
\Psi = \begin{pmatrix}
\psi_-
\\
\psi_+
\end{pmatrix},
\end{eqnarray}
 where $\psi_+$ and $\psi_-$ are (left- and right-handed) two-spinors, which may be projected out from $\Psi$ with the operators
$P_\pm = \frac{1}{2} \left(I \pm \hat{\gamma}_5 \right)$ where
\beq
\hat{\gam}^5 = i \hat{\gam}^0 \hat{\gam}^1 \hat{\gam}^2 \hat{\gam}^3 =
\begin{pmatrix} -I & O \\ O & I
\end{pmatrix} .
\eeq
We define also the Dirac conjugate
\begin{eqnarray}
\overline{\Psi} \equiv \Psi^\dagger \alpha,
\end{eqnarray}
 with $\Psi^\dagger$ denoting the usual Hermitian conjugate and 
$\alpha = -\hat{\gamma}^0$.
%The Hermitizing matrix $\alpha$ must satisfy the conditions
%\begin{eqnarray}
%\alpha \gamma^\mu + {\gamma^\mu}^\dagger \alpha = 0, \quad \quad \partial_\mu \alpha + \Gamma_\mu^\dagger \alpha + \alpha \Gamma_\mu  = 0.
%\end{eqnarray}
%We choose .

The spinor covariant derivative $\hat{D}_\nu$ 
which enters the Dirac equation (\ref{ES})
is
\begin{eqnarray}
\hat{D}_\nu =  \partial_\nu - \Gamma_\nu .
\end{eqnarray}
The spinor connection matrices $\Gamma_\nu$ are defined, up to an additive multiple of the unit matrix, by the relation
\begin{eqnarray}
\partial_\nu \gamma^\mu + \tensor{\Gamma}{^\mu _{\nu \lambda}} \gamma^\lambda - \Gamma_\nu \gam^\mu + \gam^\mu \Gamma_\nu = 0~,
\label{eq:Gamdef}
\end{eqnarray}
where $ \tensor{\Gamma}{^\mu _{\nu \lambda}}$ is the affine connection. A suitable choice satisfying (\ref{eq:Gamdef}) makes use of the spin connection $\omega_{\alpha \, b c}$,
\begin{eqnarray}
\Gamma_\alpha = -\frac{1}{4} \omega_{\alpha \, b c} \hat{\gamma}^b \hat{\gamma}^c ,
\end{eqnarray}
the spin-connection $\omega_{\mu a b}$  being defined as
\begin{eqnarray}
\tensor{\omega}{_\mu ^a _b} = e^a_\nu e^\lambda_b \tensor{\Gamma}{^\nu _{\mu \lambda}} - e_b^\lambda \partial_{\mu} e^a_{\lambda} .
\end{eqnarray}
Also, the covariant derivative of the conjugate spinor is
\begin{eqnarray}
\hat{D}_{\mu} \overline{\Psi} = \partial_\mu \overline{\Psi} + \overline{\Psi} \Gamma_\mu.
\end{eqnarray}

\setcounter{equation}{0}

%%%%%%%%%%%%%%%%%%%%%%%%%%%%%%%%%%%%%%%%%%%%%%%%%%%%%%%%%%%%%%%%%%%%%%%%%%%%%%
\section{Self-interacting Proca field: a no hair result}
%%%%%%%%%%%%%%%%%%%%%%%%%%%%%%%%%%%%%%%%%%%%%%%%%%%%%%%%%%%%%%%%%%%%%

As for  the previous study
\cite{Herdeiro:2016tmi}
 for a non-self-interacting field, the theorem is established by contradiction.
Let us assume the existence of a regular BH solution of the Einstein-Proca equations.
The general Ansatz and the field equations derived in Section 2
apply also to this case.
Differently from the globally regular case,
the geometry would possess a non-extremal horizon at, say,
$r=r_H>0$, which requires that
 \begin{equation}
N(r_H)=0 \ ,
\end{equation}
while $\sigma(r_H)>0$.
%since $r=r_H$ is a null surface.
Since we are assuming that there are no more exterior horizons,
 then $r>r_H=$constant are timelike surfaces and $N'(r_H)>0$.
 Also, we can choose without loss of generality that $\sigma(r_H)>0$,
 since the equations of motion are invariant under $\sigma\rightarrow -\sigma$.
 It follows that $N(r)$ and $\sigma(r)$ are strictly positive functions
for any $r>r_H$.

 In establishing the theorem, we shall use
the relation
	(\ref{lorentz})
	written in the generic form
\begin{eqnarray}
\label{sP1}
\frac{d}{dr}
\left(
r^2 N\sigma H\dot U
\right )
=-\frac{w r^2 F \dot U}{N \sigma},
\end{eqnarray}
together with one of the Proca equations 
\begin{eqnarray}
\label{sP2}
F'=w H \left(1-\frac{2N\sigma^2 \dot U}{w^2} \right).
\end{eqnarray}
The regularity of the horizon implies that the energy density of the Proca field
is finite there
and also the norm of the Proca-potential ${\cal A}$.
One can easily see that this condition implies
 \begin{equation}
F(r_H)=0\ .
\end{equation}
Then the function $F(r)$ starts from zero at the horizon
and remains strictly positive (or negative) for some $r$-interval
(the case of a negative $F'(r)$ can  be discussed in a similar way).
Now, let us assume $F'(r)>0$ for
$r_H<r<r_1$. 
It follows that, in this interval, $F(r)$
is a strictly increasing (and positive) function.
%As a result, the norm of the Proca-potential
%necessarily vanishes at the horizon,   ${\cal A}(r_H)=0$

Next, we consider the expression (which appears in Eqs.~(\ref{sP1}), 
(\ref{sP2})),
 \begin{equation}
P(r)\equiv 1-\frac{ 2\sigma^2(r)N(r) \dot U }{w^2}\ .
\end{equation}
One can see that $P(r_H)=1$; actually $P$
becomes negative for large $r$, since $N\to 1$, $\sigma\to 1$ as $r\to \infty$, while
 ${\cal A}^2 \to 0$
and
 $\mu>w$ 
(which is a bound state condition necessary for an exponential decay of the Proca field at infinity).
 But the important point is the existence of an $r-$interval
$r_H<r<r_2$ where $P(r)$ is a strictly positive function.
Now, the same  reasoning applies also  for $\dot U$ (since $\dot U(r_H)=\mu^2/2>0$),
while $\dot U \to 0$ asymptotically.
This implies the existence of some interval in the vicinity of the horizon where
$\dot U$, $F$ and $P$
are all positive.

At this point, let us consider an arbitrary value of $r$ in this interval.
Then
  we observe that (\ref{sP1}) implies
\begin{equation}
 r^2\sigma(r)N(r)H(r) \dot U  =-w\int_{r_H}^r\ dx \frac{x^2}{\sigma(x)N(x)} F(x) \dot U(x)<0
\label{eq21}
\end{equation}
in that interval.
Consequently, $H(r)<0$, since $\sigma,N$ are positive everywhere outside the horizon.

The last conclusion implies a contradiction: $H(r)<0$ is not compatible with $F'(r)>0$, in that interval.
 In fact, $F'(r)>0$ together with $P>0$ and $w>0$, from (\ref{sP2}), that $H(r)>0$.
Thus we conclude that $F(r)=H(r)=0$ is the only solution compatible with a BH geometry ($q.e.d.$).

%%%%%%%%%%%%%%%%%%%%%%%%%%%%%%%%%%%%%%%%%%%%%%%%%%%%%%%%%%%%%%%%%%%%%%%%%%%%%%

%%%%%%%%%%%%%%%%%%%%%%%%%%%%%%%%%%%%%%%%%%%%%%%%%%%%%%%%%%%%%%%%%%%%%%%%%%%%%%

\end{document}